\documentclass{emulateapj}
\usepackage{amsmath}
\usepackage{amssymb}
\usepackage{graphicx}
\usepackage{grffile}
\usepackage[usenames,dvipsnames]{color}
\usepackage{bm}
\usepackage{natbib}
\usepackage{hyperref}
\usepackage{balance}

\newcommand{\HI}{{\text{H\MakeUppercase{\romannumeral 1}}} }

\newcommand{\Lya}{\ifmmode{{\rm Ly}\alpha}\else Ly$\alpha$\ \fi}
\newcommand{\cm}{\ifmmode{{\rm cm}}\else cm\fi}
\newcommand{\ccm}{\,\mathrm{cm}^{-3}}
\newcommand{\ergps}{\,{\rm erg}\,{\rm s}\ifmmode{}^{-1}\else ${}^{-1}$\fi}
\newcommand{\Mpch}{\,{\rm Mpc}\,\ifmmode h^{-1}\else $h^{-1}$\fi}
\newcommand{\kms}{\,\mathrm{km}\,\mathrm{s}^{-1}}
\newcommand{\dd}{\mathrm{d}}
\newcommand{\vek}[1]{\bm{#1}}

\begin{document}
\title{Ly$\alpha$ Spectra from Multiphase Outflows, and their Connection to Shell Models}

\author{M. Gronke and M. Dijkstra}
\affil{Institute of Theoretical Astrophysics, University of Oslo, Postboks 1029 Blindern, 0315 Oslo, Norway}
\email{maxbg@astro.uio.no}

%\maketitle

\begin{abstract}
We perform Lyman-$\alpha$ (Ly$\alpha$) Monte-Carlo radiative transfer calculations on a suite of $2500$ models of multiphase, outflowing media, which are characterized by $14$ parameters. We focus on the Ly$\alpha$ spectra emerging from these media, and investigate which properties are dominant in shaping the emerging Ly$\alpha$ profile. Multiphase models give rise to a wide variety of emerging spectra, including single, double and triple peaked spectra. 
We find that the dominant parameters in shaping the spectra include (i) the cloud covering factor, $f_c$, in agreement with earlier studies, and (ii) the temperature and number density of residual HI in the hot ionized medium.
We attempt to reproduce spectra emerging from multiphase models with `shell models' which are commonly used to fit observed Ly$\alpha$ spectra, and investigate the connection between shell-model parameters and the physical parameters of the clumpy media. In shell models, the neutral hydrogen content of the shell is one of the key parameters controlling Ly$\alpha$ radiative transfer. Because Ly$\alpha$ spectra emerging from multi-phase media depend much less on the neutral hydrogen content of the clumps, the shell model parameters such as HI column density (but also shell velocity and dust content) are generally not well matched to the associated physical parameters of clumpy media. 
\end{abstract}


\section{Introduction}
The Lyman-$\alpha$ (Ly$\alpha$) emission line can be used to detect -- and study -- galaxies up to high redshifts \citep[for reviews see, e.g.,][]{Barnes2014,Dijkstra2014_review,Hayes2015}. Because \Lya is a resonant line, a \Lya photon will scatter frequently before reaching us. This implies that Ly$\alpha$ photons `sample' a wider region than merely that from where they were emitted. Ly$\alpha$ photons might thus contain unique information on the interstellar, circumgalactic and even intergalactic medium (ISM, CGM and IGM, respectively). However, it is currently unclear how well we can extract this information from observations.

The theory describing \Lya radiative transfer has been studied for decades, and for a range of gas geometries. \citet{Neufeld1990} performed an analytic study of \Lya transfer through a uniform, static semi-infinite `slab' (later transferred to spherical and cubical geometries by \citealt{Dijkstra2006} and \citealt{2006ApJ...648..762T}, respectively). \citet{Loeb1999} also presented analytic solutions for Ly$\alpha$ transfer through a homogeneous, neutral, zero-temperature intergalactic medium. Monte-Carlo codes enabled studies of Ly$\alpha$ transfer through arbitrary gas configurations. These include -- with increasing complexity -- spherically symmetric clouds \citep{Zheng2002,Dijkstra2006}, an outflowing shell \citep{Ahn2002,Verhamme2006,Schaerer2011}, conical outflows \citep{Kramer2012,Behrens2014,2014ApJ...794..116Z}, clumpy ISM models \citep{Neufeld1991,Hansen2005,Kramer2012,Laursen2012,Duval2013}, and geometries from $N$-body and/or hydro-dynamical simulations \citep[e.g.][]{Laursen07,Zheng2010,Behrens2014a,Smith2014}. 

Ly$\alpha$ transfer on interstellar scales is a complex problem which depends sensitively on the distribution and kinematics of neutral gas. Observations indicate that outflows of neutral gas promote Ly$\alpha$ escape \citep{1998A&A...334...11K,2008A&A...488..491A,Steidel2010a}. The existence of cold ($T\sim 10^4$\,K), neutral hydrogen gas in outflows with velocities of a few hundreds of km s$^{-1}$ implies that these outflows likely were multi-phase. Addressing the problem of modeling Ly$\alpha$ radiative transfer through the multi-phase ISM is beyond current models, as it would require first-principle models for both the multi-phase ISM and stellar feedback processes \citep[e.g.][]{Fujita2009}. Both these processes lie at the heart of understanding galaxy formation, and will be the subject of intense research for at least the next decade.

 \Lya radiative transfer can be represented with simplified sub-grid models for interstellar radiative transfer: the simple `shell-model'  reproduces a wide variety of observed spectra with only five parameters \citep[e.g.,][]{Verhamme2014,Martin2015,Hashimoto2015,Yang2015}. The simplicity of the shell model allows us to understand the radiative transfer process and the impact of each parameter on it in great detail. On the other hand, the shell-model is clearly an over-simplification of the problem. Especially the connection between the shell model parameters and the actual physical properties of the scattering medium is not well understood. An alternative model is the `clump' model, which consists of cold, moving clumps of neutral hydrogen embedded within a static, hot inter-clump medium (ICM). The clump model is theoretically motivated by the expectation that the ISM is multi-phase \citep{Field1969ApJ...155L.149F,McKee1977}. \citet{Neufeld1991} first introduced this model to explain Ly$\alpha$ escape from dusty media, and to show that these models may enhance the \Lya equivalent width \citep[the so-called `Neufeld effect', also see][]{Hansen2005}. This effect was later studied in more detail by \citet{Laursen2012},
\citet{Duval2013}, and \citet{Gronke2014a} who all concluded that this effect is unlikely to happen in a more-realistic environment due to the penetration of the \Lya photons in the (moving) clouds. Clumpy outflows have also been used to explain Ly$\alpha$ absorption and emission around star forming galaxies \citep{Steidel2010a,Kramer2012}, and there is increasing evidence for the existence of numerous cold, dense clumps of HI gas within the circumgalactic medium of massive galaxies \citep{Cantalupo2014,2015Sci...348..779H}.
In spite of the popularity of clump models, \Lya radiative transfer studies have thus far focused mainly on the escape of Ly$\alpha$, but not on the spectral shape\footnote{An exception to this is \citet{Duval2013} who touched upon \Lya spectral shapes as well. However, as they restricted their analyses to `clumpy shells' they only focussed on a small sub-set of clumpy outflow models.}. In this work we focus in particular on three main points:
\begin{enumerate}
\item What range of \Lya spectral shapes can we reproduce with the `clumpy' models?
\item Which of the model parameters predominantly affect the emergent \Lya spectrum?
\item Can shell models reproduce spectra obtained from clumpy models? If so, how do shell-model parameters relate to those of the clumpy model?
\end{enumerate}

The paper is structured as follows: In Sec.~\ref{sec:method} we describe our method. We present our results in Sec.~\ref{sec:results} and discuss them subsequently in Sec.~\ref{sec:discussion}. We conclude in Sec.~\ref{sec:conclusion}. Moreover, we show additional results in Appendix~\ref{sec:varying_params}.

\begin{figure}
  \centering
  \plotone{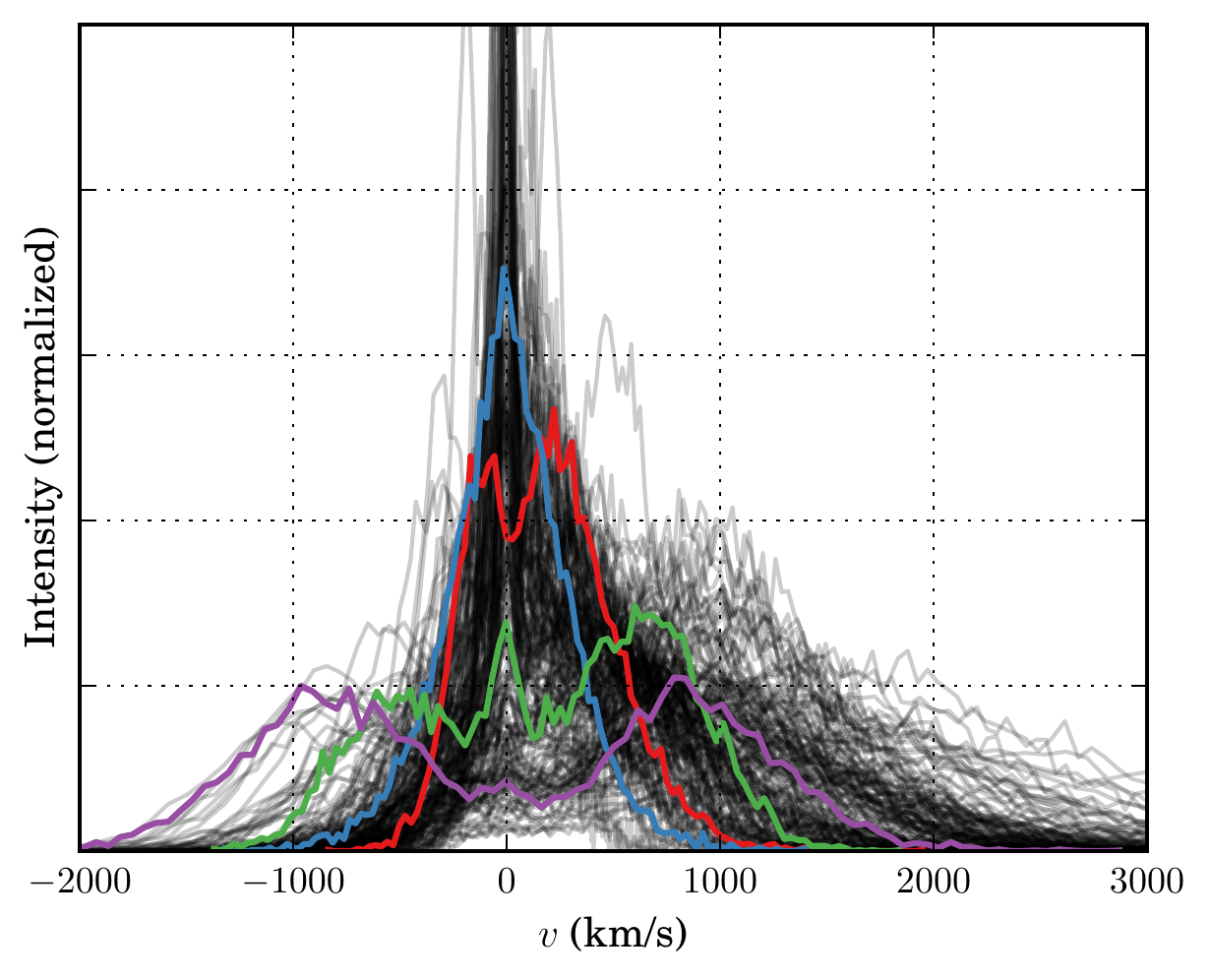}
  \caption{Examples of the range of spectra produced with the clumpy model studied.
Highlighted (in color \& thicker line) are: the `fiducial' set of parameters (see \S\,\ref{sec:clumpy-ism-model}) in red, a single peaked spectrum (in blue), a triple peaked spectrum (green), and a wide double peaked spectrum (purple). Shown in black are $200$ random spectra computed as described in \S\,\ref{sec:clumpy-ism-model}.\\}
  \label{fig:spectrum_examples}
\end{figure}

\section{Method}
\label{sec:method}

\subsection{Monte-Carlo radiative transfer}
\label{sec:monte-carlo-radi}

In this work we used the Monte-Carlo (MC) radiative transfer code \texttt{tlac} which was already used in \citet{Gronke2014a} and \citet{Gronke2015}. We refer the reader to these papers for details but here we show the basics of the calculation.

In a nutshell, MC radiative transfer simulations work by following the trajectories of individual photon packages in space- and frequency-space and, thus, slowly converging towards the solution. The relevant steps included are 
\begin{enumerate}
\item the \textit{emission} of photons, where the photon is assigned an initial random direction and frequency according to some probability density function (PDF);
\item the \textit{propagation} of a distance $d$, where $\tau = \int_0^d (n_{\HI}\sigma_{\HI} + n_d\sigma_d)\dd s$ is a random variable drawn from an exponential PDF. In the above equation $n_\HI$ ($n_d$) and $\sigma_\HI$ denotes the number density and cross section of neutral hydrogen (dust), respectively.
\item the \textit{interaction} of the photon with a particle where either a new frequency and direction are assigned, or, the photon is absorbed.
\item the \textit{output} of the photons' relevant quantities once it has escaped the simulation box.
\end{enumerate}

One difference to the approach in \citet{Gronke2014a} is the use of a partial core-skipping scheme. In particular, we do not use core skipping in the ICM ($x_{\rm crit} = 0$) and use a the dynamical core-skipping \citep[similar to the technique used by][]{Smith2014} inside the clumps\footnote{Core-skipping is an acceleration technique which works by forcing the photon into the wing of the line by drawing the scattered atom's velocity from a truncated Gaussian with cutoff $x_{\rm crit}$. Inside the clumps, we use $x_{\rm crit}  =  0$ for $a \tau_{0, \rm esc} < 1$ and $x_{\rm crit} = (a \tau_{0, \rm esc})^{1/3} / 5$ otherwise. Here, $a=4.7\times 10^{-4} (T / 10^4{\rm K})^{-1/2}$ denotes the Voigt parameter and $\tau_{0, \rm esc} = n_{\HI} \sigma_{\HI}(x = 0) (r_{\rm cl} - r)$ where $r$ is the distance from the photon's position to the cloud center and $r_{\rm cl}$ is the cloud radius.}. We verified that the results are indistinguishable from the runs without non-core skipping.

\begin{table*}
  \centering
  \caption{Overview of the model parameters.}
\begin{tabular}{l|lrrr}
\hline
  Parameter  & Description & Fiducial value & Allowed range & Units \\ \hline\hline
$v_{\infty,\,{\rm cl}}$ & Radial cloud velocity & $100.0$ & [$0.0$, $800.0$] & $\,{\rm km\,s}^{-1}$ \\
$r_{\rm {\rm cl}}$ & Cloud radius & $100.0$ & [$30.0$, $200.0$] & $\,{\rm pc}$ \\
$P_{\rm {\rm cl}}$ & Probability to be emitted in cloud & $0.35$ & [$0.0$, $1.0$] &  \\
$H_{\rm em}$ & Emission scale radius & $1000.0$ & [$500.0$, $3.0\times 10^{3}$] & $\,{\rm pc}$ \\
$f_{\rm {\rm cl}}$ & Cloud covering factor & $3.5$ & [$0.8$, $8.0$] &  \\
$T_{\rm {\rm ICM}}$${}^{\dagger}$ & Temperature of ICM & $10^{6}$ & [$3.0\times 10^{5}$, $5.0\times 10^{7}$] & $\,{\rm K}$ \\
$n_{\rm HI,\,{\rm ICM}}$${}^{\dagger}$ & \HI number density in ICM & $5.0\times 10^{-8}$ & [$10^{-12}$, $10^{-6}$] & $\,{\rm cm}^{-3}$ \\
$\sigma_{\rm i}$ & Width of emission profile & $50.0$ & [$5.0$, $100.0$] & $\,{\rm km\,s}^{-1}$ \\
$T_{\rm {\rm cl}}$${}^{\dagger}$ & Temperature in clouds & $10^{4}$ & [$5.0\times 10^{3}$, $5.0\times 10^{4}$] & $\,{\rm K}$ \\
$\beta_{\rm cl}$ & Steepness of the radial velocity profile & $1.5$ & [$1.1$, $2.5$] &  \\
$\tilde\sigma_{\rm d, cl}$${}^{\dagger}$ & Dust content in clumps & $3.2\times10^{-22}$ & [$4.7\times 10^{-24}$, $1.6\times 10^{-21}$] & $\,\cm^{2}$ \\
$\zeta_d$${}^{\dagger}$ & Ratio of ICM to cloud dust abundance & $0.01$ & [$10^{-4}$, $0.1$] &  \\
$\sigma_{\rm {\rm cl}}$ & Random cloud motion &$40.0$ & [$5.0$, $100.0$] & $\,{\rm km\,s}^{-1}$ \\
$n_{\rm HI,\,{\rm cl}}$${}^{\dagger}$ & \HI number density in clouds & $0.35$ & [$0.03$, $3.0$] & $\,{\rm cm}^{-3}$ \\
\hline
\end{tabular}
\tablecomments{Variables marked with ${}^{\dagger}$ were drawn in log-space (see \S\,\ref{sec:clumpy-ism-model}).}

\vspace{.5cm}
  \label{tab:models}
\end{table*}

\subsection{The clumpy ISM model}
\label{sec:clumpy-ism-model}

The clumpy ISM parametrization used in this work is adapted from \citet{Laursen2012} and already used in \citet{Gronke2014a}. We therefore refrain from describing the parameters again in great detail. A brief overview over the parameters is given below.
\begin{itemize}
\item The geometry of the setup is described by the radius of the simulation sphere $r_{\rm gal}=5\,$kpc, the cloud radius $r_{\rm cl}$, and the covering factor $f_{\rm cl}$ which is the number of clouds on average passed from the center.

\item The content of the cold [hot] clumps [inter-clump medium (short: ICM)] is given by $T_{cl},\, n_{\HI, cl}$ [$T_{ICM},\, n_{\HI, ICM}$] for temperature\footnote{The temperature given can be seen as an effective temperature with the effects of turbulence included.} and the number density of hydrogen, respectively. We express the effect of dust in terms of dust cross section per hydrogen atom, i.e., $\tilde\sigma_d\equiv \tau_d / (n_\HI d)$ where $\tau_d$ is the dust optical depth and $d$ the path-length considered\footnote{Using simplifying assumptions, this can be related to the metallicity via $\tilde\sigma_d = Z/Z_\sun \sigma_d$ where $\sigma_d\approx 1.58\times 10^{-21} \cm^{2}$ \citep{Pei1992,Laursen2009}}. Furthermore, we choose to parametrize the dust content in the ICM as $\zeta_d \equiv \tilde\sigma_{d, {\rm ICM}}/ \tilde\sigma_{d, {\rm cl}}$. These number densities lead to a theoretical column density which a photon has to cross before escaping the simulation box given by
\begin{equation}
N_\HI = n_{\HI, {\rm ICM}} (r_{\rm gal} - H_{\rm em}) + \frac{4}{3} \tilde f_{\rm cl} r_{\rm cl} ( n_{\HI, {\rm cl}} - n_{\rm ICM})
\label{eq:N_HI}
\end{equation}
where $\tilde f_{\rm cl} = f_{\rm cl}\frac{r_{\rm gal} - H_{\rm em}}{r_{\rm gal}}$ is the reduced covering factor which reflects that \Lya photons are emitted throughout the cloud, rather than in its center\footnote{One might argue that because of the isotropic initial direction of the photons one should replace $r_{\rm gal} - H_{\rm em}$ by merely $r_{\rm gal}$ in the above expressions. However, as \Lya turns to escape via lower-column densities we choose to use Eq.~\eqref{eq:N_HI} -- which can be seen as a lower limit of a hydrogen column density assigned to a clumpy ISM} and $H_{\rm em}$ is the scale length of the emission radius as described below.

\item The emission properties of the photon are controlled with $\sigma_i$ and $P_{\rm cl}$ which give the intrinsic width of the line and the probability that a photon is emitted within a cloud, respectively. The emission site is drawn randomly from $r \sim \exp(-r / H_{\rm em})$ where $r$ is the distance to the center of the simulation domain.

\item In addition, the clouds move with \textit{(i)} an isotropic velocity component with magnitude drawn from a Gaussian with standard deviation $\sigma_{\rm cl}$ (which is equivalent to draw each Cartesian component from a Gaussian with $\sigma=\sigma_{\rm cl} / \sqrt{3}$), and \textit{(ii)} a radial velocity component given by 
\begin{equation}
  v(r) = v_{\infty,{\rm cl}}\left\{1 - \left(\frac{r}{r_{\text{min}}}\right)^{1-\beta_{\rm cl}}\right\}^{1/2}
\label{eq:velocity_profile}
\end{equation}
 for $r>r_{\text{min}} = 1\,{\rm kpc}$ and otherwise zero.
\end{itemize}

This leaves us with $14$ free parameters\footnote{Note, that the parameters given here differ slightly from what we used in \citet{Gronke2014a}. There, we ignored the filling of the ICM since we were interested in the (enhancement of) the \Lya escape fraction.} which are listed in Table~\ref{tab:models}. There, also our fiducial parameters are given which define our \textit{fiducial model}. We chose these values to be centered on what \citet{Laursen2012} calls ``realistic parameters'' with the exception of the outflow velocity $v_{\infty, {\rm cl}}$ where \citet{Laursen2012} chose deliberately small values.

Equipped with this model parametrization we assembled a library of $2,500$ spectra (using $\sim 10,000$ \textit{escaped} photons each). We drew each parameter uniformly withing the allowed range listed in Table~\ref{tab:models} which is loosely based on the ``extreme'' range of \citet{Laursen2012}. Note, that $n_{\HI, {\rm ICM}},\,n_{\rm d, ICM},\,T_{\rm ICM},\,T_{\rm cl},\,\tilde\sigma_{d, \rm cl}$ and $\zeta_d$ were drawn in log-space (marked with ${}^\dagger$ in Table~\ref{tab:models}).

\begin{figure}
  \centering
\plotone{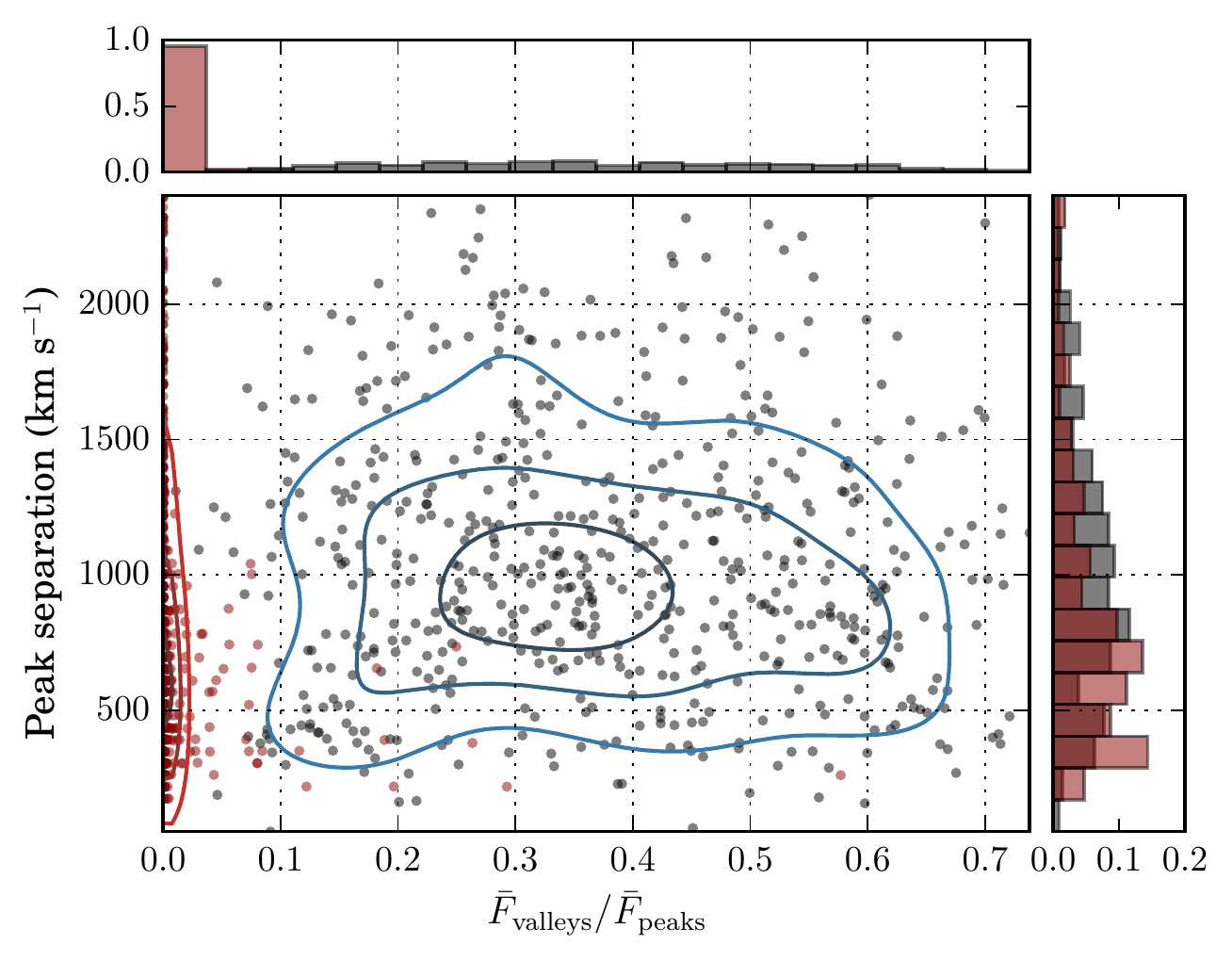}
  \caption{Maximal peak separation of double-peaked spectra versus the ratio of the means fluxes at the peaks and the valley for the clumpy model spectra (grey points, blue contours) and shell model spectra (red points \& contours). The contours show the $(0.5,\,1.0,\,1.5)\sigma$ levels calculated with a Gaussian kernel density estimator. See \S\,\ref{sec:spectra} for details. \\}
  \label{fig:scatter1}
\end{figure}
\begin{figure}
  \centering
  \includegraphics[width=\linewidth]{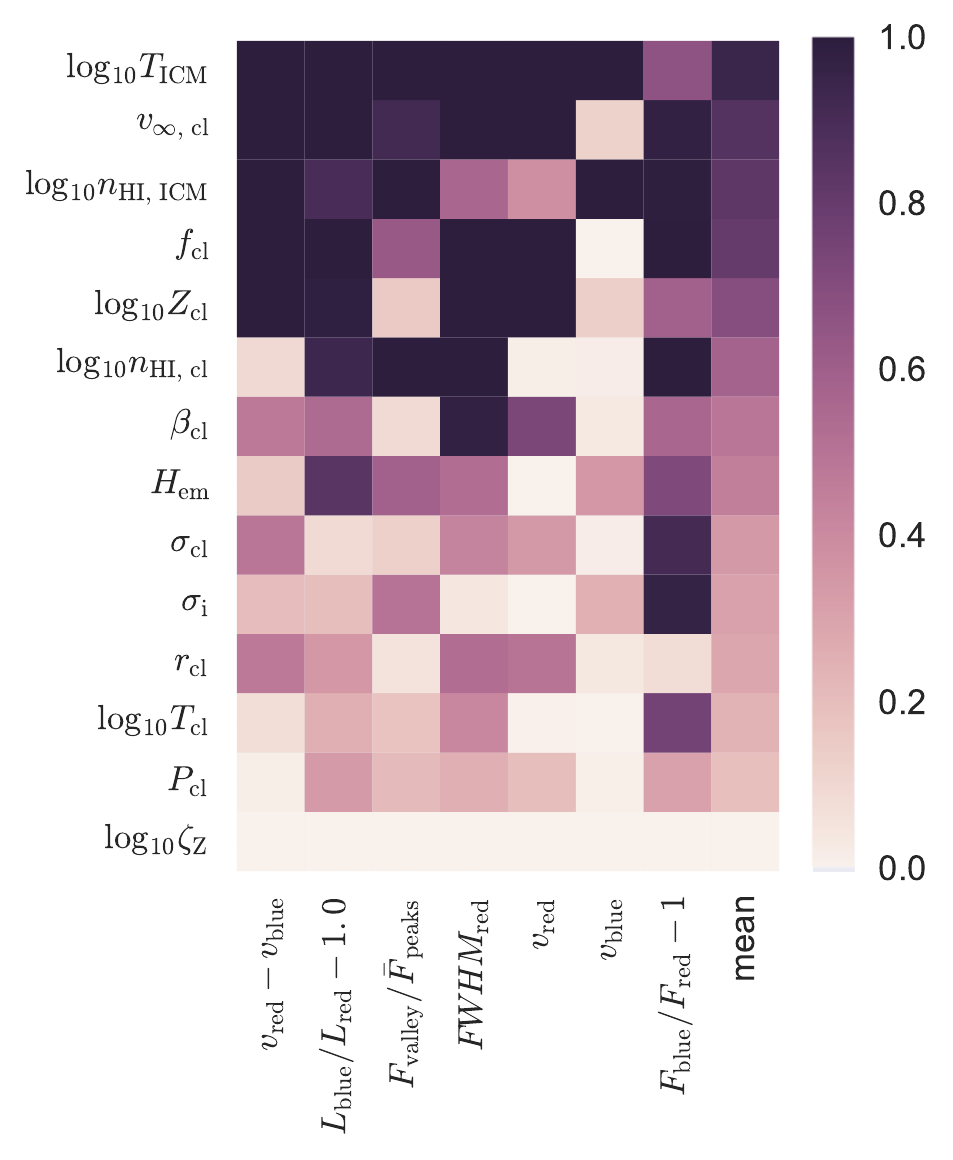}
  \caption{Relevance score of the model parameters fitting the spectral features of the double peaked spectra. The scores were computed using `lasso' analysis as described in \S\,\ref{sec:sens-spectr-shap} and reflect how strongly a parameter affects an observable.\\}
  \label{fig:linear_lasso}
\end{figure}

\begin{figure*}
  \centering
  \includegraphics[width=.95\textwidth]{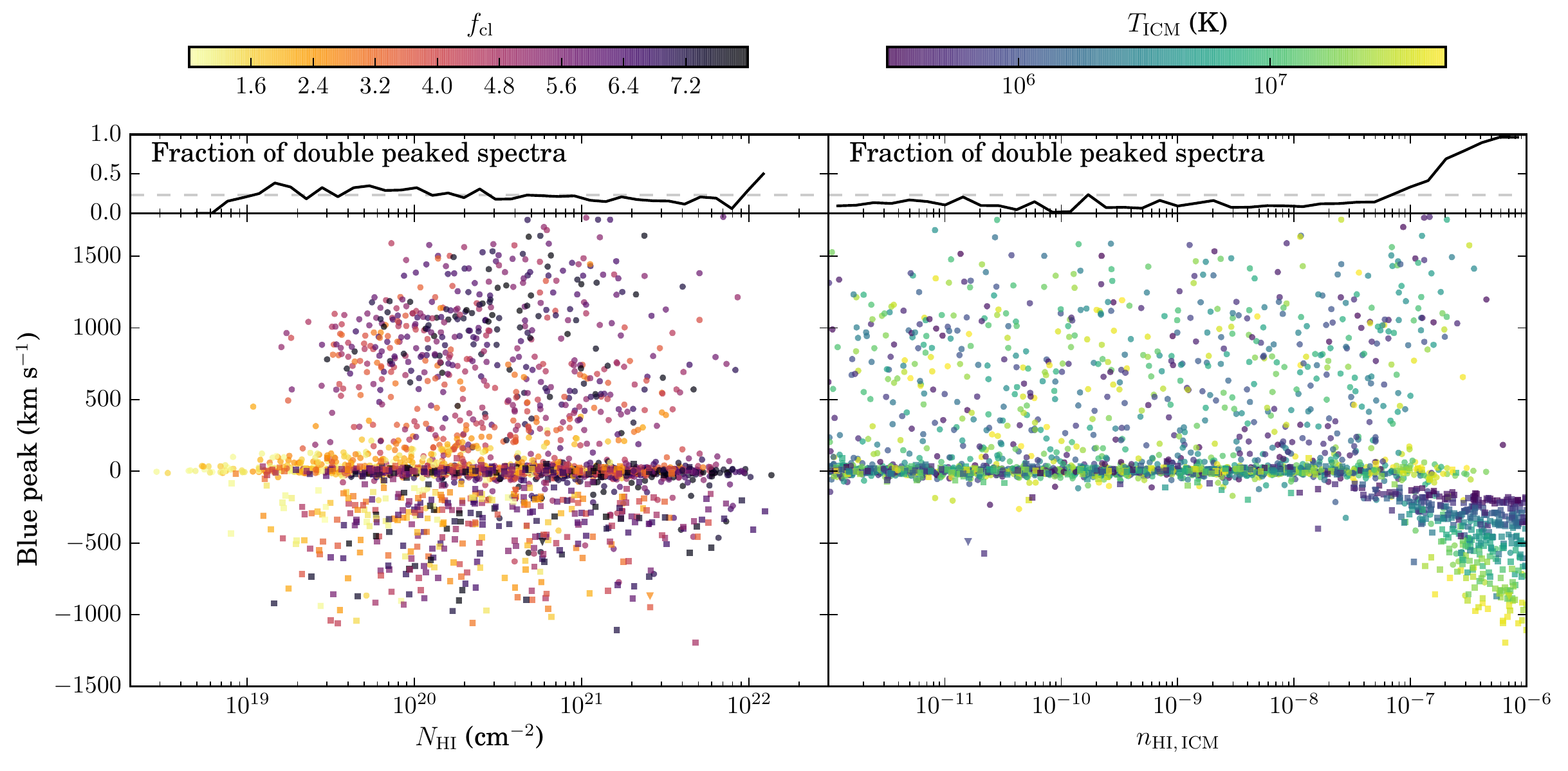}
  \caption{The position of the blue-most peak versus the total column density \textit{(left panel)} and vs the number density of the ICM \textit{(right panel)}. In addition the colors denote the value of $f_{\rm cl}$ and $T_{{\rm ICM}}$, respectively. Circles, squares, triangles represent single-, double- and triple-peaked spectra, respectively. 
\\}
  \label{fig:scatter2}
\end{figure*}

\subsection{Spectra characterization}
\label{sec:spectra-char}
The simulation output is binned using the \citet{FDrule} rule to optimize the relative error per bin without being too sensitive to outliers. We assigned to each bin the conservative error of $1.3$ times the Poisson error with a minimum error in each bin of the mean error of the Poisson error of the spectrum. This choice can be interpreted as adding a systematic (instrumental) error to the spectrum and ensures that the fitting procedure (see \S\,\ref{sec:shell-model-fitting}) is not controlled by individual bins with a very small error (e.g., with no flux), and widens the set of `good fits' as well as the uncertainty on the model parameters. As both turn out to be quite small (see \S\,\ref{sec:shell-model-fits}) we think the additional error is not overestimated.

The resulting \Lya spectra are classified as single-, double- or triple-peaked spectra (we did not find any spectra with more than three peaks in our data set) using a peak detecting algorithm. The algorithm\footnote{We used a modified version of \url{https://gist.github.com/sixtenbe/1178136} which takes the error of the spectrum correctly into account.} flags a peak (a valley) if the following $N_{\rm pd}$ data points are at least a value of $\delta_{\rm pd}$ times the error in this region smaller (greater) than the candidate. For our purpose, we executed the algorithm for $N_{\rm pd} = (1,\,2,\,3,...,\,10)$ with the final result being the median of the detected number of peaks (under the constraint that the number of valleys found is one less than the number of peaks found). This procedure -- which we verified visually using $\sim 10\%$ of the spectra -- ensures a robust characterization.

Once the peaks and valleys of a given spectrum are detected, the spectral shape can be characterized by a number of extracted parameters. Although a unique characterization does not exist, we will adapt the commonly used quantities $F_{\rm blue}$ ($F_{\rm red}$) and $FWHM_{\rm blue}$ ($FWHM_{\rm red}$) for the maximum flux and the full width at half maximum of the blue-most (red-most) peak, respectively. Furthermore, we analyze $v_{\rm blue}$ ($v_{\rm red}$) and $L_{\rm blue}$ ($L_{\rm red}$) which defines the position and integrated flux of these peaks.

\subsection{Shell model fitting}
\label{sec:shell-model-fitting}
The shell-model commonly used in \Lya radiative transfer studies consists of a central, luminous source surrounded by an outflowing shell of hydrogen and dust. This geometry is described by five parameters: the width of the intrinsic emission line $\sigma_i$, the outflow velocity $v_{\rm exp}$, the (effective) temperature $T$ (which includes a Doppler component due to turbulence), and the hydrogen column density and dust optical depth of the shell $N_\HI$ and $\tau_d$, respectively. In order not to confuse these parameters with the ones from the clumpy model (presented in \S\,\ref{sec:clumpy-ism-model}), we denote the shell model parameters with a superscript ${}^{\rm sm}$ or ${}^{\rm shell\ model}$.

The shell model has been used to fit observed \Lya spectra \citep[e.g., recently by ][]{Verhamme2014,Yang2015,Patricio2015} with surprising success -- given the simplicity of the model compared to the complexity of galactic structures. In the following, we use the improved pipeline described by \citet{Gronke2015} to fit shell model spectra to \Lya spectra obtained from the clumpy model. In short, the systematic pipeline consists of a library of $12,960$ spectra covering the three discrete parameters $T$, $N_{\HI}$ and $v_{\rm exp}$ whereas the other parameters are modelled by weighting each photon individually \citep[see][for details]{Gronke2015}. In order to fit a spectra, two steps are performed:  \textit{(i)} the global maximum is found using a basinhopping algorithm \citep{basinhopping1} -- where we use the \citet{scipy} implementation, and \textit{(ii)} to find the uncertainties as well as the degeneracies between parameters around that maximum, we use the affine-invariant Monte-Carlo sampler \texttt{emcee}\footnote{In order to be able to better sample on a multi-modal landscape we also used the parallel tempered sampler. However, since the local maxima differ usually by $\log p \gtrsim 5$, we returned to the vanilla sampler.} \citep{Goodman2010,Foreman-Mackey2012}.

Compared to the version described in \citet{Gronke2015}, we improved the fitting pipeline in two ways. First, step \textit{(i)} is now performed on the discrete and continuous parameters in an alternating fashion (i.e. the basin-hopping algorithm on the discrete parameters calls a second basin-hopping algorithm on the continuous ones). Due to the possibility to use more advanced local minimizing algorithms on the continuous parameters, this enables a faster and more robust global maximum finding procedure. This is important as the overall likelihood has several disjoint local maxima and a wrongly chosen starting position will not be resolved during the sampling process. 
The second improvement is the extension of the parameter space to negative outflow velocities, and to lower column densities down to $N_{\rm HI}=10^{16}\,\cm^{-2}$. We achieved the negative outflow velocities by flipping the photons' frequencies around line center, as done previously by \citet{Schaerer2011}. The extension of the grid of discrete models was done to allow for lower optical depth solutions as they are more likely in a clumpy ISM\footnote{Note, that we omitted the `core-skipping' technique for models with $a(T) N_{\HI} \sigma_{\HI}(v = 0, T) < 200$ and re-computed the models required.}.

Except otherwise stated, we terminate the basin-hopping after $100$ iterations (on the discrete grid) of unchanged results, and run the Monte-Carlo algorithm with $700$ steps and $400$ walkers.

\section{Results}
\label{sec:results}

\subsection{Characteristics of spectra}
\label{sec:spectra}
Varying the parameters described in \S\,\ref{sec:clumpy-ism-model} gives rise to a wide variety of \Lya spectra. Fig. \ref{fig:spectrum_examples} visualizes this fact by superimposing four particular spectra (in color) over $200$ randomly drawn spectra. The four selected spectra feature the double peaked fiducial spectrum (in red), a single peaked (blue), an asymmetric triple peaked spectrum (in green), and a wide double peaked profile (in purple).

From our sample of $\sim 2500$ spectra, we found that $\sim 77\%$ are single, $\sim 23\%$ double peaked, and $\lesssim 1\%$ triple peaked. We caution, however, that the characterization can be very difficult in some cases.
Even within the subset of double-peaked profiles there is a big variation in the considered spectral parameters (see \S\,\ref{sec:spectra-char}). For example, the peak separation $v_{\rm red} - v_{\rm blue}$ has a value of $1007.70^{+596.33}_{-488.24}\,\kms$ (where we show the median and the difference to the $16$th and $84$th percentile to illustrate the asymmetric distribution). As expected from an outflow, the red peak is more pronounced than the blue peak which is quantified with the found ratios of the integrated fluxes $L_{\rm blue} / L_{\rm red} - 1 = -0.63^{-0.20}_{+0.40}$ and the maximum fluxes $F_{\rm blue} / F_{\rm red} - 1 = -0.63^{-0.20}_{+0.40}$. In addition, the location of the red peak is further away from line center than its blue counterpart ($792.32^{-457.62}_{+402.87}\kms$ versus $-218.86^{-335.05}_{+224.18}\kms$). This shift is also present in the single-peaked spectra where the peak position is $15.39^{-30.12}_{+740.22}\kms$.

Fig.~\ref{fig:scatter1} focuses on an interesting correlation of double peaked spectra, namely between the peak separation and the ratio of the mean fluxes of peaks and valleys. The one- and two-dimensional projections of the distribution reveal that clumpy models fill a large portion of the plane spanned by the two observables. This is particularly interesting as double peaked shell-model spectra are mostly restricted to $F_{\rm valley} \lesssim 0.1 \bar F_{\rm peaks}$. We obtained this result by drawing $1000$ random shell-model spectra from the parameter range described in \S\,\ref{sec:shell-model-fitting} and show their features in Fig.~\ref{fig:scatter1} as comparison.
This difference might be useful to distinguish between the two models (also see \S\,\ref{sec:shell-model-fits}).

\begin{figure*}
  \centering
  \plotone{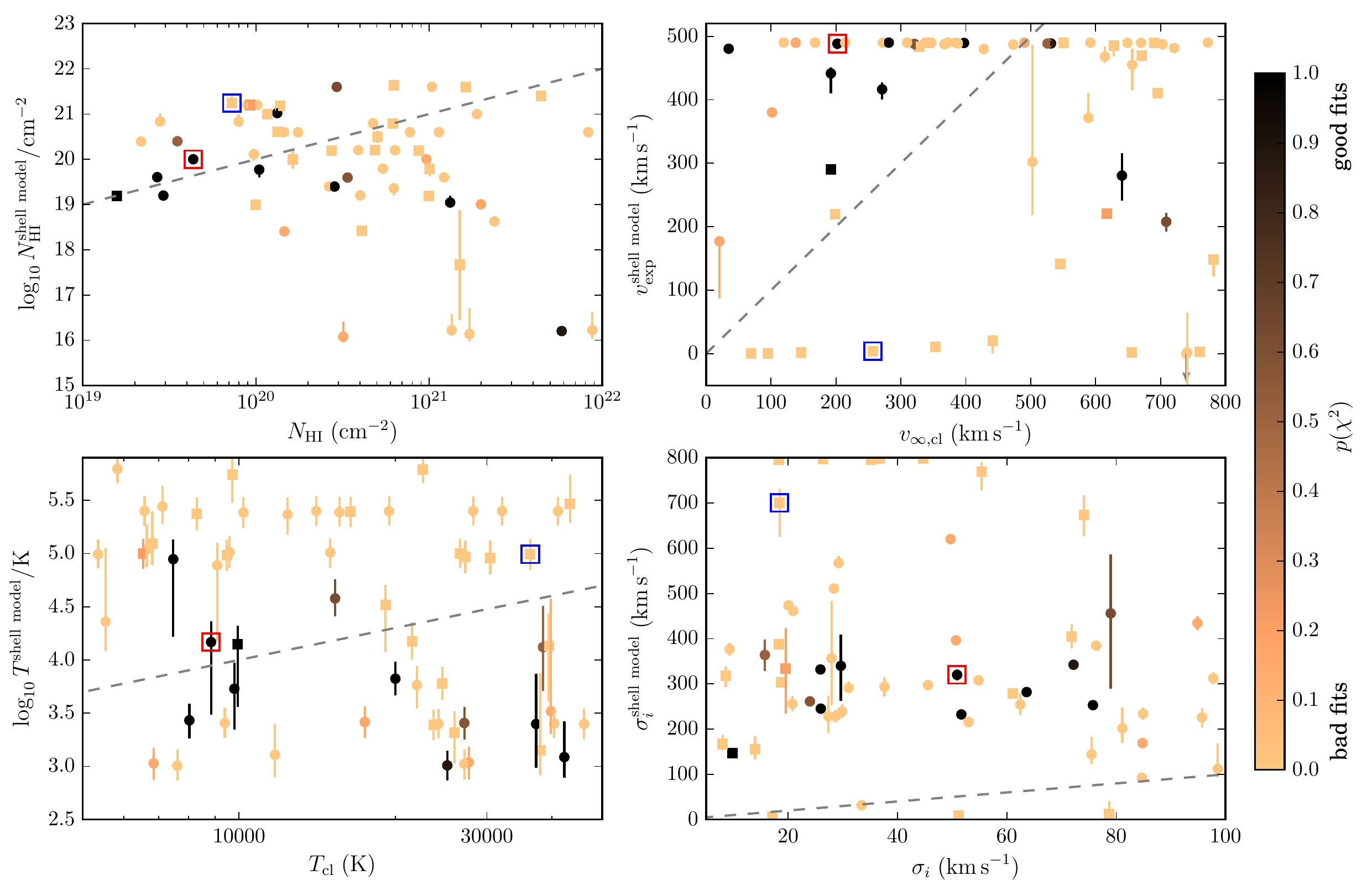}
  \caption{Results of the shell model fitting as described in \S\,\ref{sec:shell-model-fits}. On the $x$ axis the (clumpy model) input parameters are given and on the $y$ axis the recovered shell-model parameters. The marker and errorbars denote the $50$th, $16$th and $84$th percentile, respectively. Additionally, circles, squares and triangles represent single-, double- and triple-peaked spectra, and the color shows the quality of the fit. The grey dashed line marks the one-to-one relation \& the grey arrows indicate if a data point lies outside the range displayed. Note, that the spectra of the points surrounded by squares are shown in Fig.~\ref{fig:sm_fitspecs}. \\}
  \label{fig:sm_multiplot}
\end{figure*}

\subsection{Sensitivity of spectral shapes to multiphase medium }
\label{sec:sens-spectr-shap}

In order to analyze what correlations exist between the model parameters and the spectral shape parameters we performed a `lasso' analysis \citep{tibshirani1996regression} using the \texttt{python} library \texttt{skikit-learn} \citep{scikit-learn}\footnote{The `lasso' works by adding to the usual linear minimization $1/(2 N)\sqrt{\sum_{i=0}^{N} (y_i - \vek{\hat x}\cdot\vek{x}_i)^2}$ a penalty of $\alpha |\vek{\hat x}|_1$, where $|\vek{\hat x}|_1$ denotes the L1 norm of the vector $\vek{\hat x}$. This leads to the sparsity of $\vek{\hat x}$, i.e., the reduction of relevant parameters. This analysis is repeated for a number of values for $\alpha$ and different sub-samples of the data. The measure of `relevance' is then the fraction of times a parameter was included in $\vek{\hat x}$.}. Fig.~\ref{fig:linear_lasso} shows the summarized results of the `lasso' analysis with the darker color corresponding to a more relevant parameter for a given observable. This Figure shows, for example, that $f_{\rm cl}$, $T_{\rm ICM}$, and $n_{\rm ICM}$ are the parameters that affect the emerging spectrum most strongly. In contrast, $n_{\HI, {\rm cl}}$ has considerably less impact on the spectra.
Generally, we find a (much) larger impact of the ICM parameters on the spectral shape and not the cloud parameters. We will elaborate on this in \S\,\ref{sec:sens-lya-spectr}.\footnote{In addition, we also performed a second order analysis which includes also the products of each parameter with the other ones (results not shown). We found that under the top parameters also the cross terms between the temperatures and hydrogen number densities appear.}

Figure~\ref{fig:scatter2} shows examples of how parameters affect certain spectra characteristics. In particular we plotted the position of the blue peak of the spectrum versus a certain model parameter. Note, that for single peaked spectra (marked with circles), the position of the blue peak is identical with the position of the only peak. In addition, we show the fraction of double peaked spectra above scatter plot. 
In the left panel of Fig. \ref{fig:scatter2} no correlation can be seen between the total column density (see Eq.~\ref{eq:N_HI}) and the position of the blue peak as well as the spectral type. In this panel we color-coded the value of $f_{\rm cl}$ which affects $N_{\rm HI}$ most strongly.
For comparison, the right panel shows an example of a nice correlation between the position of the blue peak and the content of the ICM. Clearly, a hotter ICM leads to a shift of the blue peak if $n_{\HI, {\rm ICM}}\gtrsim 3\times 10^{-8}\ccm$ as well as an tremendous increase in double peaked spectra. This shows visually the great relevance of $\log T_{\rm ICM}$, $n_{\HI, {\rm ICM}}$ and also the mixed term $\log T_{\rm ICM}\times n_{\HI, {\rm ICM}}$.

\subsection{Shell model fits}
\label{sec:shell-model-fits}
Fig.~\ref{fig:sm_multiplot} shows the results of the shell-model fitting (described in \S\,\ref{sec:shell-model-fitting}) for $60$ randomly chosen spectra (out of which $20$ double-peaked, $40$ single-peaked spectra). In the figure, the marker symbol stands for the spectral type, the color of the marker illustrates the quality of the fit, and in each panel we plot the recovered shell model parameter (with its uncertainties) against a particular clumpy model parameter. Overall, we found roughly a third of the spectra reasonably well recovered ($p(\chi^2) > 0.01$, in $18$ cases). 

In the upper left panel of Fig.~\ref{fig:sm_multiplot} we show the recovered shell-model column density against the calculated column density of the clumpy model (see Eq.~\ref{eq:N_HI}). This shows that the correct column density is generally not recovered -- independent of the type of spectrum or the quality of the fit. On average, $N_{\rm HI}$ is underestimated by $~0.5-1$ orders of magnitude. For individual spectra, however, the offset can be as large as $3\,$dex, and in either direction. Only in $\sim 15\%$ of the cases the correct value was recovered.

The upper right panel of Fig.~\ref{fig:sm_multiplot} shows the recovered expansion velocity of the shell versus the $v_{\infty, {\rm cl}}$ parameter (see Eq.~\ref{eq:velocity_profile}). As $v_{\rm exp}^{\rm sm}$ is a fixed velocity for all the hydrogen in the system whereas $v_{\infty, {\rm cl}}$ is the limiting bulk cloud velocity, we expected to find $v_{\rm exp}^{\rm sm}\lesssim v_{\infty, {\rm cl}}$. This is also the case for some of the analyzed spectra where we found values of $v_{\rm exp}^{\rm sm}\sim 0\kms$. Interestingly, this group consists only of double-peaked spectra 
However, another big fraction of the spectra preferred very high values of $v_{\rm exp}^{\rm sm}\sim 500\kms$. Here, $500\kms$ is also the limit of our grid of shell-models (see \S\,\ref{sec:shell-model-fitting}), and therefore we can expect even larger values for the best-fit values of $v_{\rm exp}^{\rm sm}$. In this group of models, we find single- as well as double-peaked profiles with input parameters covering the whole allowed range of $v_{\infty, {\rm cl}}$. Also, the fit quality is wide-spread -- independently of the spectral type or the particular value of $v_{\infty, {\rm cl}}$.

The lower row of Fig.~\ref{fig:sm_multiplot} shows the temperature and the intrinsic spectral widths relations (in the left and right panel, respectively). As commonly known \citep[see, e.g.,][]{Schaerer2011,Gronke2015}, the shell-model spectral shapes are generally not very sensitive to the (effective) temperature of the shell. Hence, the large uncertainties on $T^{\rm sm}$. The added value of the recovered temperatures and $\sigma_i$ is somewhat questionable as the geometries and emission sites are quite different. It is, however, interesting that \textit{(i)} no clear clustering or correlation is visible, i.e., the spectral types and fit qualities seem to be well mixed, and \textit{(ii)} the values of $\sigma_i$ are generally overestimated by a factor of a few. Especially, the latter point can be understood as emission within (randomly moving) clouds leads to an effective broadening of the intrinsic spectrum. 

\begin{figure*}
  \centering
  \plotone{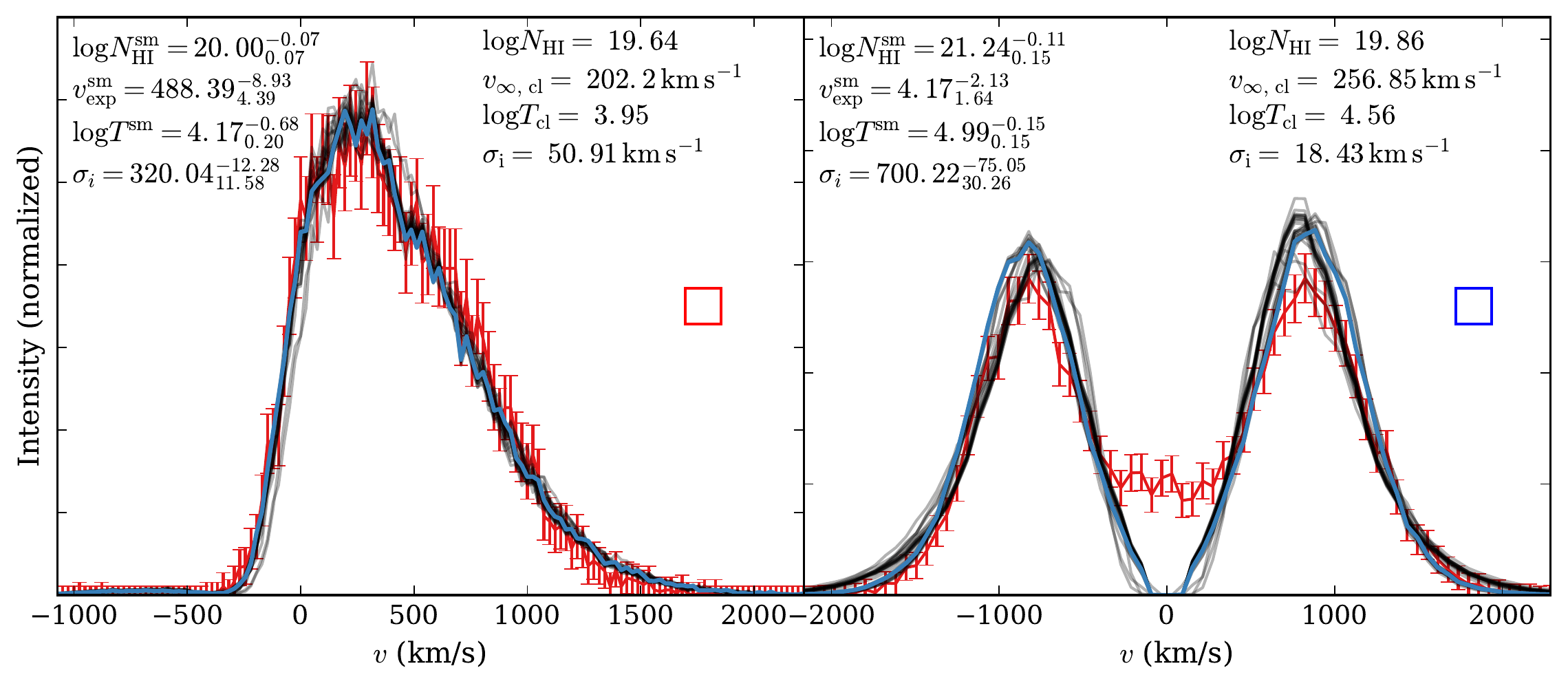}
  \caption{Two examples of the fitting results. The \textit{red points} show the spectra of a clumpy model with associated uncertainty (see \S\,\ref{sec:spectra-char}). The \textit{black solid lines} show 25 random spectra from the burned-in MC chain, and the best-fit spectrum is drawn with a \textit{blue solid line}. In the upper left corner the $16$th, $50$th, and $84$th percent quantiles of the shell-model parameters are displayed. As a comparison their clumpy models `counterparts' are shown in the upper right corner. Note, that we highlighted the two spectra shown here in Fig.~\ref{fig:sm_multiplot} using a red and blue rectangle.\\}
  \label{fig:sm_fitspecs}
\end{figure*}

The red and blue rectangles drawn in Fig.~\ref{fig:sm_multiplot} highlight examples which are shown in Fig.~\ref{fig:sm_fitspecs}. In particular, Fig.~\ref{fig:sm_fitspecs} shows the clumpy model spectra in red, and the best-fit shell model spectra in blue. In order to illustrate the uncertainty of the shell-model fits, we also display $25$ spectra which are randomly drawn from the burnt-in Monte-Carlo, and thus distributed as given by the sampled likelihood function. We choose the two spectra because they help us illustrate particular problems with shell model fitting (see \S\,\ref{sec:unus-shell-model}). \\

\section{Discussion}
\label{sec:discussion}

In this section, we discuss the impact of our results. In particular, we focus on the set of spectral shapes achievable through our clumpy model parametrization in \S\,\ref{sec:sens-lya-spectr}, comment on the connection to the shell-model in \S\,\,\ref{sec:unus-shell-model}, and bridge to the observational side in \S\,\ref{sec:comp-observ} and \S\,\ref{sec:impact-cgm}.

\subsection{Sensitivity of the \Lya spectrum to the clump \& ICM properties}
\label{sec:sens-lya-spectr}
Our result show that the emergent \Lya spectrum is more sensitive to the ICM than the cloud parameters. To first order, this is already visible when comparing the left and right panels of Fig.~\ref{fig:temperatures} and Fig.~\ref{fig:hydrogen} in the appendix where we varied the temperatures and hydrogen contents in the clouds and ICM separately. 
The same result becomes apparent in the more detailed analysis presented in \S\,\ref{sec:sens-spectr-shap} (see Fig.~\ref{fig:linear_lasso} \& \ref{fig:scatter2}): $n_{\HI, {\rm ICM}}$ and $T_{\rm ICM}$ play a much more important role than $n_{\HI, {\rm cl}}$ and $T_{\rm cl}$, respectively. One should note, however, that this is only true for $n_{\HI, {\rm ICM}}\gtrsim 3\times 10^{-8}\ccm$ as Fig.~\ref{fig:scatter2} shows.
This critical value of $n_{\rm HI, ICM}$ reflects the extent of the system ($5$\,kpc) and the allowed ICM temperature range (see Table~\ref{tab:models}), which translate approximately to a optical depth at line center of $\tau_0\approx 0.4 - 5$, i.e., where the ICM is becomes optically thick for line-center photons.
Naturally, this critical value depends also on other parameters, such as $f_{\rm cl}$ or the cloud motion. Interestingly, this value lies within the `realistic' parameters of \citet{Laursen2012}, and corresponds closely to $n_{\HI}$ of the hot ionized medium in the \citet{McKee1977} picture (assuming collisional ionization equilibrium)\footnote{As the assumed HIM temperature of \citet{McKee1977} is on the lower limit of our allowed range one might conclude that we mostly overestimated the ICM hydrogen number density. However, two factors should be taken into account \textit{(i)} the ISM might be dominated by gas which is out of equilibrium \citep[e.g.][]{Walch2015}, and \textit{(ii)} temperatures are effective temperatures with the small-scale turbulence included whereas the ionization relies on the absolute temperature.} 

Two notable exceptions to this rule are the cloud covering factor $f_{\rm cl}$ and the dust content of the clouds $\tilde\sigma_{\rm d, cl}$. The former plays a major -- if not the most important -- role in determining the spectral shape.

The parameter $f_{\rm cl}$ enhances the total ICM column density encountered by \Lya photons in the ICM as $\tilde N_{\rm HI, ICM}\sim r_{\rm gal} n_{\rm HI, ICM} (f_{\rm cl} + 4/5)$ \citep[see][]{Hansen2005}. In addition, movement of the clouds (random motions and/or ordered outflowing motions) provide an efficient way of transferring Lya photons into the wings of the line profiles. We found this second effect to be most important by varying $f_{\rm cl}$ from $0.8$ to $8$ in a static setup (i.e., $v_{\infty, {\rm cl}} = \sigma_{\rm cl} = 0$) and found -- in stark contrast to the left panel of Fig.~\ref{fig:geometry} -- only minor changes in the spectral shape\footnote{Note, that the mere movement of the clouds is needed not the structured movement. I.e., in a second test where only $v_{\infty, {\rm cl}}=0$ but $\sigma_{\rm cl}$ was left at its fiducial value the covering factor did affect the spectrum.}.

The dust content of the clumps is closely tied to the importance of $f_{\rm cl}$. An increased value of $\tilde\sigma_{\rm d,cl}$ shifts the weight of the emergent spectrum towards the photons which did not experience many scattering events within clumps. Therefore, increasing $\tilde\sigma_{\rm d,cl}$ mimics a decrease in the effective covering factor. This is apparent when comparing the right panel of Fig.~\ref{fig:dust}, and the left panel of Fig.~\ref{fig:geometry} where the increase of $\tilde\sigma_{\rm d, cl}$ as well as the decrease of $f_{\rm cl}$ lead to the disappearance of the extended red wing of the spectrum.

The result that the ICM plays such a strong role in shaping the emerging spectrum may be surprising as it apparently supports the model of \citet{Neufeld1991} and \citet{Hansen2005}. In their picture, \Lya photons `reflect off' the surface of the clumps which naturally minimizes the exposure to HI inside the cold clumps, and maximizes the exposure to HI in the ICM. \citet{Laursen2012} refuted this model as in more realistic environment (mainly with moving clouds) the \Lya photons penetrate deeper into the clouds, scatter there many times and potentially get destroyed. However, for the emergent spectrum this does not play a role as the scatterings in the ICM put the photon's frequency far into the wing (in the clouds' frame). Reversely, many scatterings in the cloud leave the photon still in the core (in the ICM frame).
This effect can be illustrated by comparing the hydrogen column densities experienced by \Lya photons $\hat N_{\HI}$ of two models with $n_{\HI, {\rm cl}}=0.1\ccm$ and $1\ccm$ (and otherwise fiducial parameters) which are $\log_{10}\left(\hat N_{\HI} / \cm^{-2}\right) = 20.3^{-0.7}_{+0.4}$ and $20.9^{-1.3}_{+0.6}$ (using the notation of \S\,\ref{sec:spectra}), respectively. Also, the number of clouds intercepted are about the same ($12^{-7}_{+10}$ versus $11^{-6}_{+9}$), supporting our explanation.

\begin{figure*}
  \centering
  \plotone{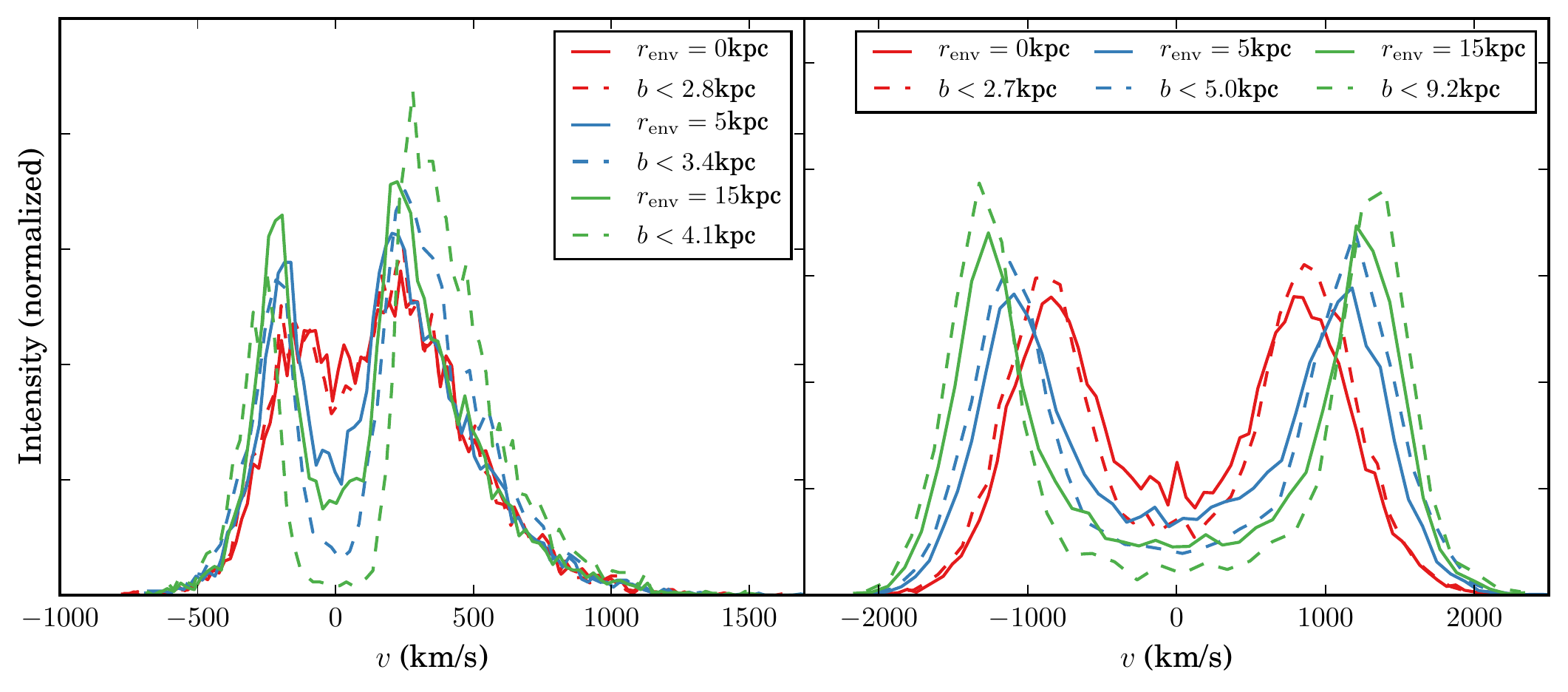}
  \caption{Impact of the environment and measurement apparatus on the emergent spectrum. The \textit{solid lines} show the variation of the spectrum with the addition of a surrounding halo (using the ICM parameters), and the corresponding \textit{dashed lines} show the spectrum with only the innermost $50\%$ of the escaping flux. As examples, we show in the left panel the fiducial spectrum (see Table~\ref{tab:models}), and in the right panel the double-peaked spectrum of Fig.~\ref{fig:sm_fitspecs}. See \S\,\ref{sec:impact-cgm} for further details.\\}
  \label{fig:environment}
\end{figure*}

\subsection{The (un)usability of shell-model fitting}
\label{sec:unus-shell-model}
Section \ref{sec:shell-model-fits} shows that the shell-model can reproduce some of the spectra of the clumpy-model, but generally cannot be used to infer the physical parameters of the host system as the inferred shell-model parameters can be several orders of magnitudes off -- independent of the quality of the fit.

The tension between $N_{\HI}$ and $N_{\HI}^{\rm sm}$ cannot be resolved if the latter is instead compared to the median of the actually experienced hydrogen column density $\hat N_\HI$, the analytically computed `experienced ICM column density' $\tilde N_{\HI, {\rm ICM}} = n_{\HI, {\rm ICM}} r_{\rm gal} (f_{\rm cl} + 4/5)$\footnote{We denote the total, actual experienced column density by the photons (obtained from the radiative transfer simulation) with $\hat N_{\rm HI}$ which should not be confused with the column density of the system $N_{\HI}$ (see Eq.~\eqref{eq:N_HI}) or the analytically computed `experienced ICM column density' for which we use $\tilde N_{\rm HI, ICM}$.}, or only the ICM column density $N_{\HI, {\rm ICM}} = n_{\HI, {\rm ICM}} r_{\rm gal}$. The reason for these mismatches is illustrated well by the simplified model discussed in \S\,\ref{sec:sens-lya-spectr}: the hydrogen column density experienced by the photons (or, the one that is given by the geometry) is dominated by the dense, cold, neutral medium, i.e., the clumps. However, the spectral shape is predominantly given by the (few) scattering events occurring in the hot, ionized medium (the ICM). This is in sharp contrast with the shell models which contains only one column density.

An immediate implication of this is the difference in the minimal flux between the peaks for double peaked profile. Fig.~\ref{fig:scatter1} shows that this is $\sim 0$ for most shell models while clumpy models allow -- due to escaping photons which experienced very little neutral hydrogen -- much greater values. As the right panel of Fig.~\ref{fig:sm_fitspecs} shows, this is one of the reasons why the shell-model cannot reproduce the symmetric, double peaked profiles of the clumpy model.

Photons that encountered low column densities of HI can escape close to the line center (or even at line center). \citet{Behrens2014} already highlighted the importance of low-column-density channels for escape of \Lya of galaxies \citep[see also  fig.~4 of][for a simple demonstration of this effect]{CR7_paper}. As these low-density escape routes -- which govern the resulting shape of the \Lya spectrum -- do not exist within the shell-model it is maybe not surprising that the actual properties (e.g., the column density) of the host system and the inferred ones from shell-model fitting do not match. Hence, one has to be cautious when assigning physical meaning to the unconverted shell-model parameters.

On the other hand, the power of the shell-model fitting should not be underestimated. As has been shown in numerous studies \citep[e.g.,][]{Verhamme2012,Yang2015} the shell-model can fit the majority of \Lya spectra observed -- with relatively few free parameters to an astonishing accuracy. 
We have demonstrated clearly that there is no simple conversion from shell model parameters to parameters describing the clumpy medium. The fact that shell models are so succesful at reproducing data suggests that there may still be a connection, but at a more subtle level. We will explore this in future studies.
However, it is already clear now (as we laid out in \S\,\ref{sec:sens-lya-spectr}) that this conversion must not ignore the crucially important hot ionized medium.

\subsection{Comparison to observations}
\label{sec:comp-observ}
As already pointed out several times in this paper, the shell-model can reproduce observed \Lya spectra. In this work, we have shown that the shell-model can also reproduce a subset of the clumpy-model spectra. However, whether or not the clumpy model can reproduce observed spectra is still an outstanding question. Currently, we can only conjecture that this is the case -- given the amount of free parameters and the established non-empty intersection with the set of shell-model spectra. 

\citet{Yang2015} used the shell-model to fit the \Lya spectra of the `Green peas' (high-redshift analogues at $z\sim 0.1-0.3$). They could reproduce nine out of twelve spectra where the remaining three have a valley position redward of line center. The obtained expansion velocities (column densities) range from $\sim 0-350\kms$ ($\sim 10^{19}-10^{20}\cm^{-2}$).
The shell-model fitting results of \citet{Hashimoto2015} ($12$ \Lya emitters at $z\sim 2.2$) show expansion velocities of $\sim 100-200\kms$ and column densities of $\log_{10}N_\HI / \cm^{-2} \sim 16 - 20$. These recent results -- with high quality \Lya spectra -- confirm, on the one hand, the ability of the shell-model to model observed spectral profiles remarkably well. This confirms the need to understand shell-model fitting in a broader picture. On the other hand, the spectra presented in \citet{Yang2015} which cannot be reproduced by the shell-model might hint towards the need for an extension of the modelling parameter space. Also, the shell model has difficulties in reproducing surface-brightness profiles of spatially extended \Lya sources (\citealt{Barnes2010MNRAS.403..870B}, but see \citealt{Patricio2015}).

A completely different interpretation of \Lya transfer was recently presented by \citet{Hagen2015}, who interpreted their detection of $12$ \Lya emitting galaxies (out of a sample of $63$) in terms of a \Lya `opening angle'. In their model, this is the combined solid angle of holes in the ISM through which \Lya can escape easily. At first glance, this interpretation can be connected easily to clumpy outflows as the opening angle can be related\footnote{Assuming a Poisson distribution of the number of clumps per sightline with mean $f_{\rm cl}$.} to our properties via $\Omega_{\Lya} \approx 4 \pi (1-P_{\rm cl}) \exp\left(-f_{\rm cl}\right)$. However, we caution that the directional dependence of \Lya escape is only weak: the directional dependence is set mostly by the last scattering event prior to escape, which causes the emerging radiation from clumpy models to be quite isotropic \citep[see][]{Gronke2014a}.

The fact that several observed \Lya spectra can be reproduced using shell-model spectra, and we found that the overlap between the studied clumpy-model and the shell-model is not very big, suggests that our current clumpy model parametrization is not sufficient to capture the full set of observer \Lya spectra. Possible extensions would be the introduction of a non-static ICM, temperature and density gradients in the ICM as well as in the clumps, a modification of the velocity profile, the introduction of one or several other phases, and the consideration of the galactic environment and the instruments (see \S\,\ref{sec:impact-cgm} for the latter point).
To conclude, the unification of the radiative transfer models and observations is still an outstanding issue.

\subsection{Impact of the galactic environment \& instruments}
\label{sec:impact-cgm}

As the setup we presented in this work represents a dusty multiphase medium the path of actual observed \Lya photons differ in two main aspects from the spectra we simulated: \textit{(i)} the photons have to pass through the immediate surrounding and the IGM before reaching us, and \textit{(ii)} not all the escaping flux of a galaxy is actually observed. 

The impact of the IGM on the spectra has been studied by several groups in the past \citep[see, e.g.,][]{Dijkstra2007a,Zheng2010ApJ...716..574Z,Laursen2011} with the conclusion that the blue side (up to $v\lesssim 100\kms$) of \Lya spectra can be strongly affected at redshifts $\gtrsim 4$. This, however, does not explain the different spectral shapes observed for lower redshifts (see \S\,\ref{sec:comp-observ}).

In contrast to the IGM, the impact of the CGM on \Lya spectra has not been studied systematically, yet. This is partly due to fact the structure and kinematics of the CGM are highly complex, and not fully understood. Observations show that 
neutral hydrogen can be found out to $\sim 300$kpc for all galaxy types \citep{Prochaska2011ApJ...740...91P}, and at least part of the CGM is in a multiphase state \citep{Steidel2010a}. This picture is also supported by state-of-the art hydro-dynamical simulations \citep[e.g.,][]{Shen2012} and allows us to put our work into the following context: \textit{(a)} our simplistic parametrized multiphase medium can be seen as a sub-grid model for the ISM as well as (at least part) of the CGM, and \textit{(b)} possible further processing of the computed \Lya spectra might be necessary before comparing it to observations. 

To mimic the effects of the galactic environment, we consider -- as a first, crude approximation -- a low-density, \HI halo out to radius $r_{\rm gal} + r_{\rm env}$ filled with the same content as the ICM.
Fig.~\ref{fig:environment} shows the changed fiducial spectrum (left panel) and double peaked spectrum from Fig.~\ref{fig:sm_fitspecs} (right panel) using these considerations. In this figure, the solid lines show the spectrum using all the emergent photons and the dashed lines in corresponding color show the spectra using only the photons within a certain impact parameter. Here, the cutoff was chosen so that half of the total escaped photons is used in the spectra. Note that the actual fraction varies and depends on \textit{(i)} the redshift, \textit{(ii)} physical properties of the object, and \textit{(iii)} on the instrument used \citep{Steidel2010a,Wisotzki2015,Momose2015}.
Fig. \ref{fig:environment} shows clearly, that the galactic environment as well as the impact parameter cutoff have a (strong) effect on the spectrum. Possibly most interesting is the apparent suppression of the flux at line center.
This might be the key to the reconciliation of the clumpy model and the observed spectra. However, the environment considered here is over-simplified as in reality we expect density and  temperature gradients \citep{2015MNRAS.448..895S,Pallottini2014} as well as coherent gas motion \citep{Bird2015}. We leave this interesting topic to future studies.\\

\section{Conclusion}
\label{sec:conclusion}

We present a systematic study of \Lya spectra emerging from simplified models of multiphase outflows. While these models are well-motivated and have been used in \Lya radiative transfer studies before, spectra from these models have barely been analyzed.\\

Our main findings are as follows:
\begin{itemize}
\item Clumpy outflows give rise to a wide range of \Lya spectra, including spectra with high flux at line center which are encountered less frequently with shell models (\S\,\ref{sec:spectra}).
\item We demonstrate that in clumpy outflows, the key parameters that predominantly determine the emerging spectra are, the covering factor $f_{cl}$ of clumps, the number density of \HI in the hot, inter-clump medium, and the temperature of the inter-clump medium. Interestingly, the radiative transfer process is less sensitive to the hydrogen contents of the clumps. This result contrasts with the shell models where the average column density of the system is one of the most important parameters regulating outcome (\S\,\ref{sec:sens-spectr-shap}, \S\,\ref{sec:sens-lya-spectr}).
\item We fit shell models to a sub-set of out clumpy models and find that generally, the  parameters of the best-fit shell models barely correlate with the physical parameters of the clumpy models (\S\,\ref{sec:shell-model-fits}, \S\,\ref{sec:unus-shell-model}).
\item Shell models can fit only a small sub-set of clumpy outflow spectra well, which is partly because clumpy models allow for much more efficient escape of \Lya photons at line center. These models agree better with the data if additional scattering in the CGM is invoked (\S\,\ref{sec:impact-cgm}).
\end{itemize}
This suggests that extracting physical information from shell model parameter is less straightforward than previously thought. We therefore caution against overinterpreting the shell model parameter, until their physical meaning is understood better. We will be addressing this in future work.

\acknowledgements
We thank D. Neufeld, L. Mas-Ribas, J. Rhoads and H. Yang for helpful discussions. MG thanks
the Physics \& Astronomy department at the Johns Hopkins University for their hospitality. MD thanks the astronomy department at UCSB for their kind hospitality.

\bibliography{references_all}

\appendix
\section{Impact of individual parameters on the spectral shape}
\label{sec:varying_params}
\begin{figure}
  \centering
  \plottwo{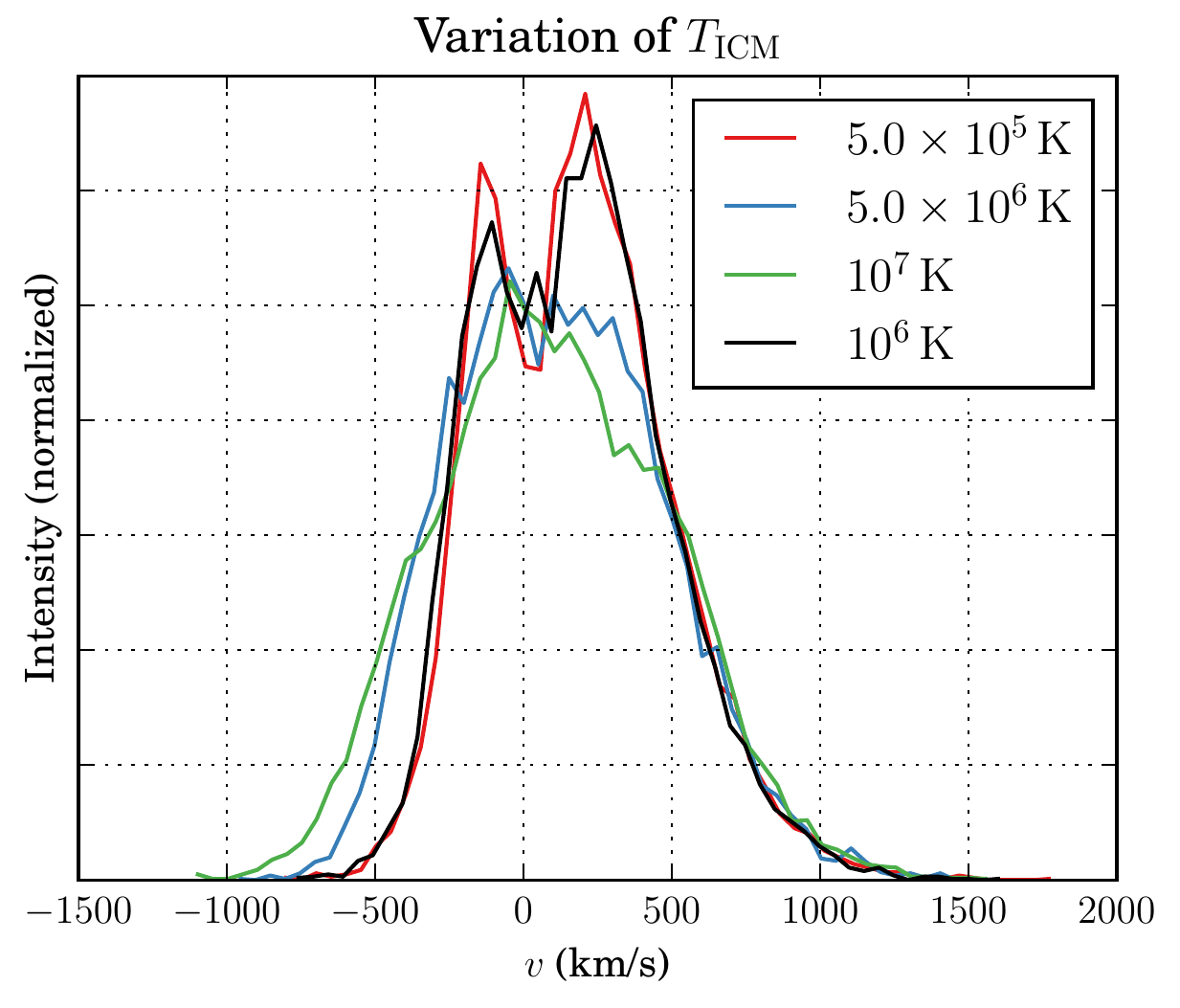}{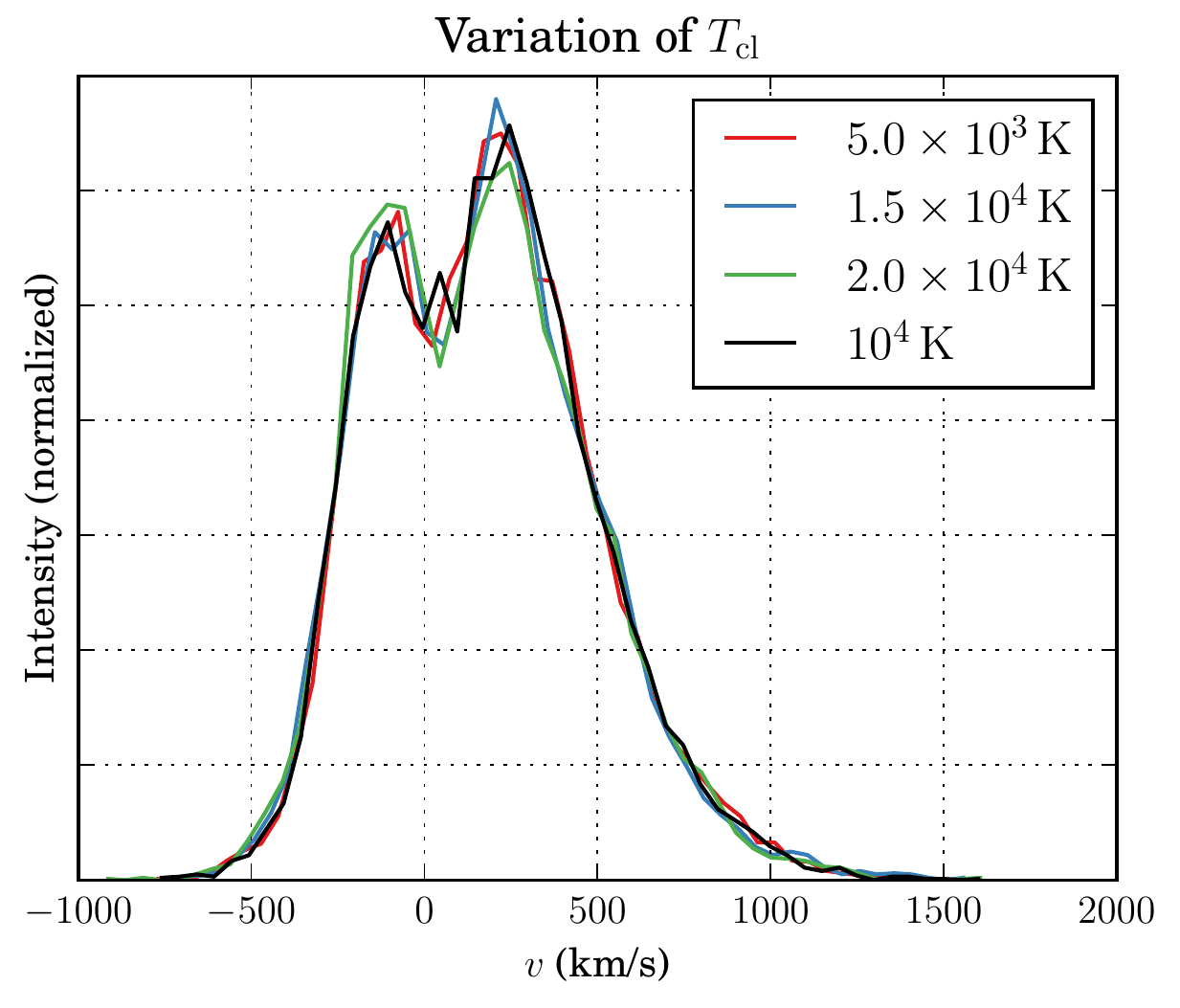}
  \caption{Impact of the gas temperature in the ICM \textit{(left panel)} and in the clouds \textit{(right panel)}.}
  \label{fig:temperatures}
\end{figure}

\begin{figure}
  \centering
  \plottwo{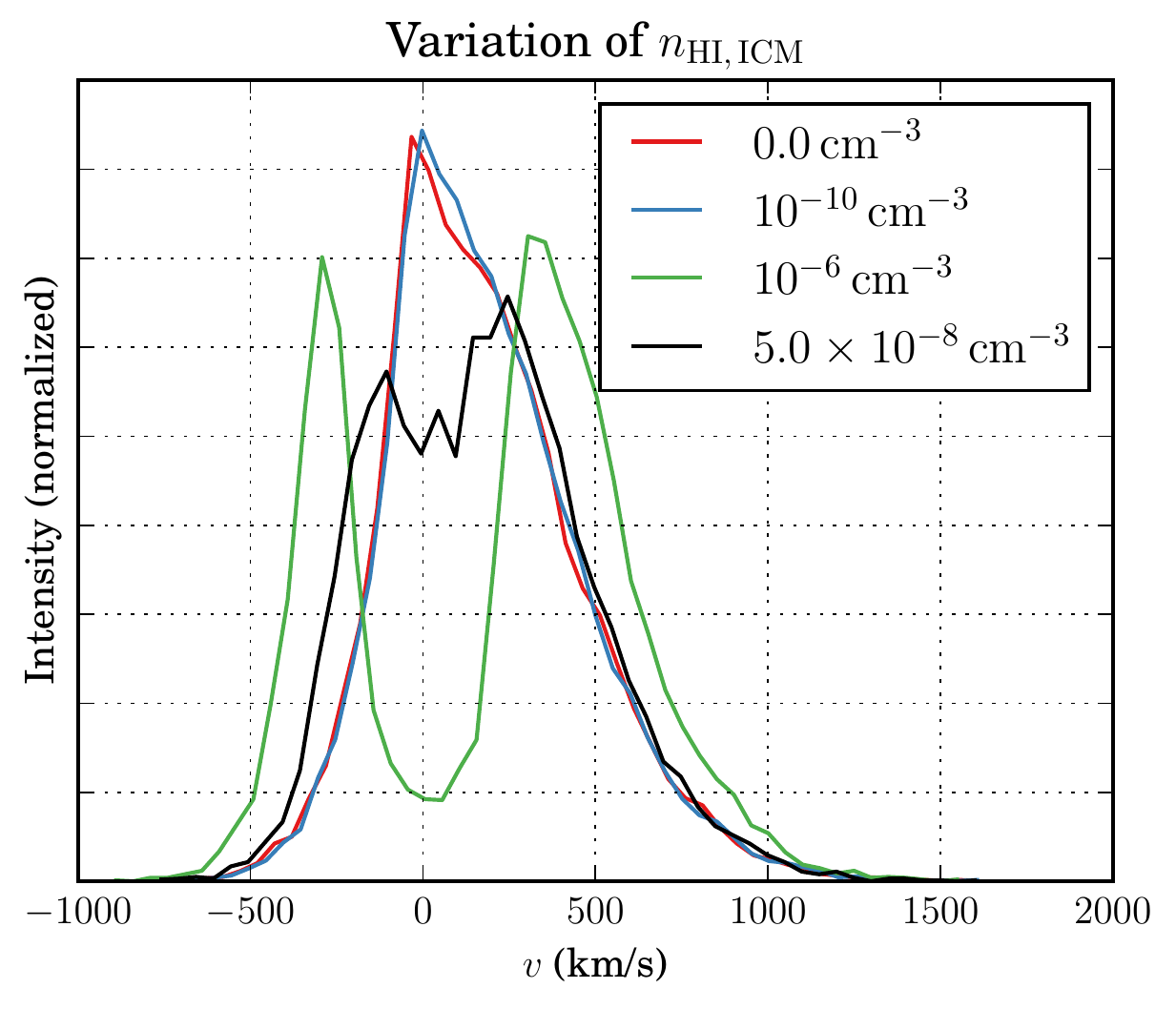}{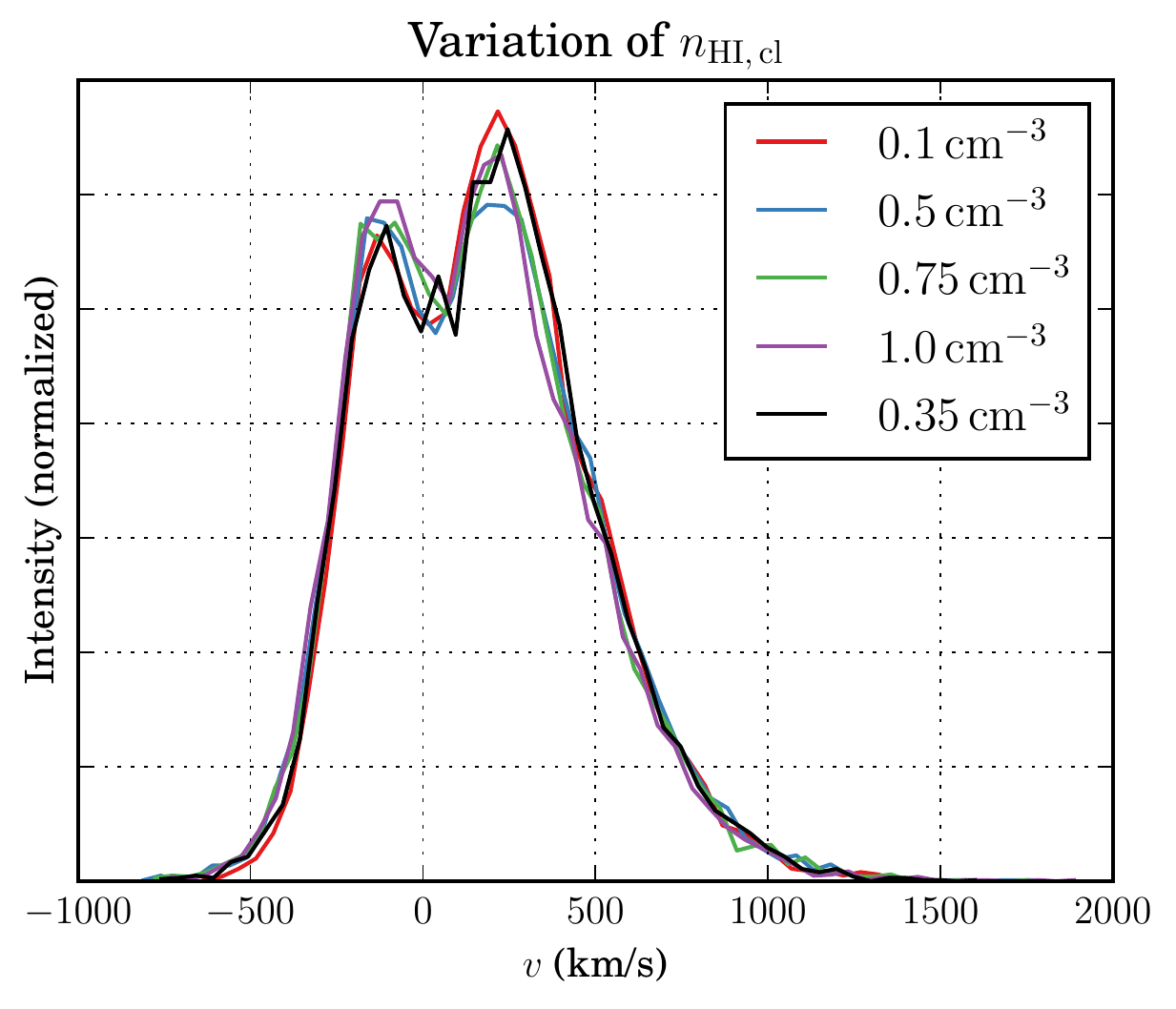}
  \caption{Variation of the hydrogen number density in the ICM \textit{(left panel)} and in the clouds \textit{(right panel)}.}
  \label{fig:hydrogen}
\end{figure}

\begin{figure}
  \centering
  \plottwo{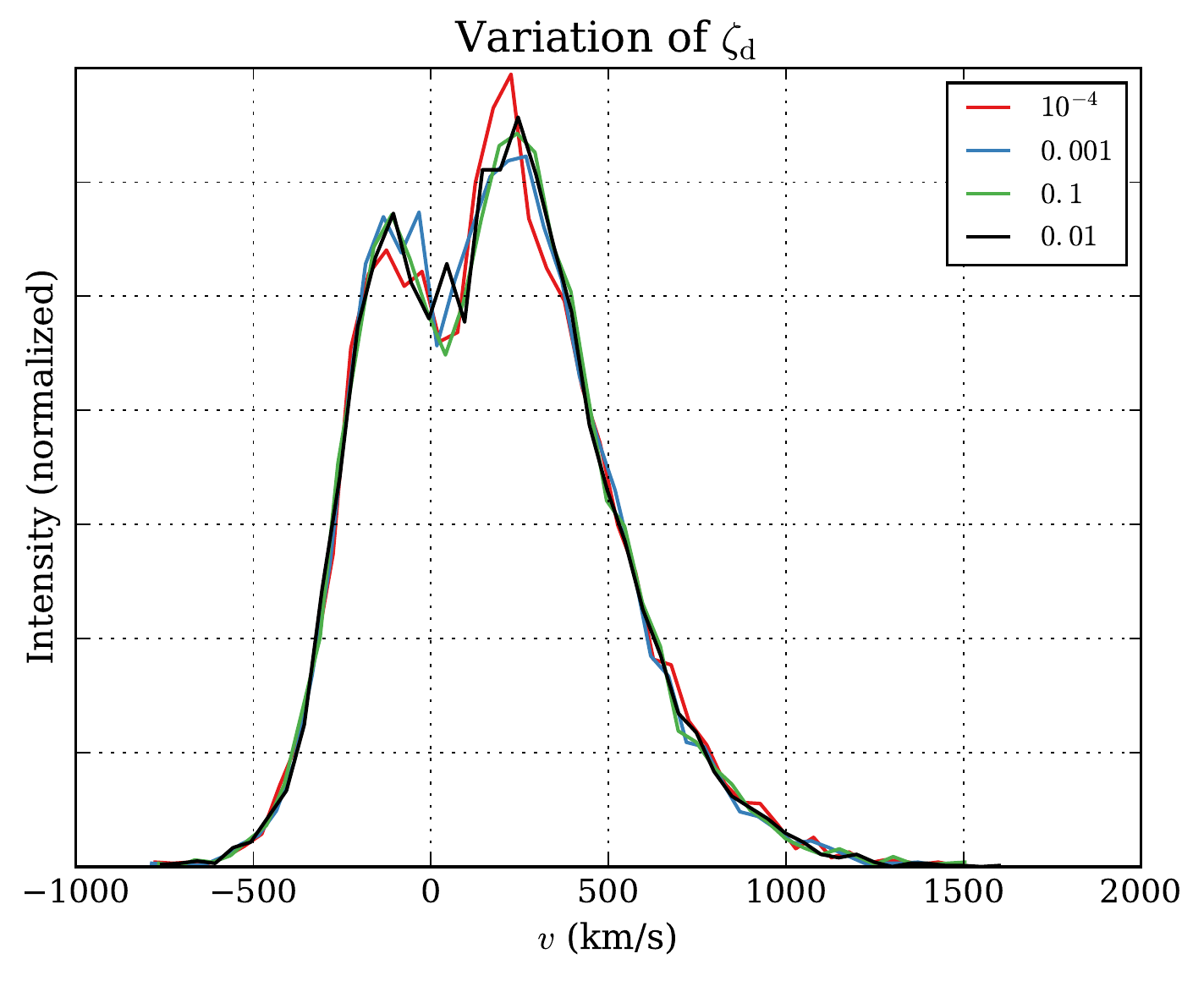}{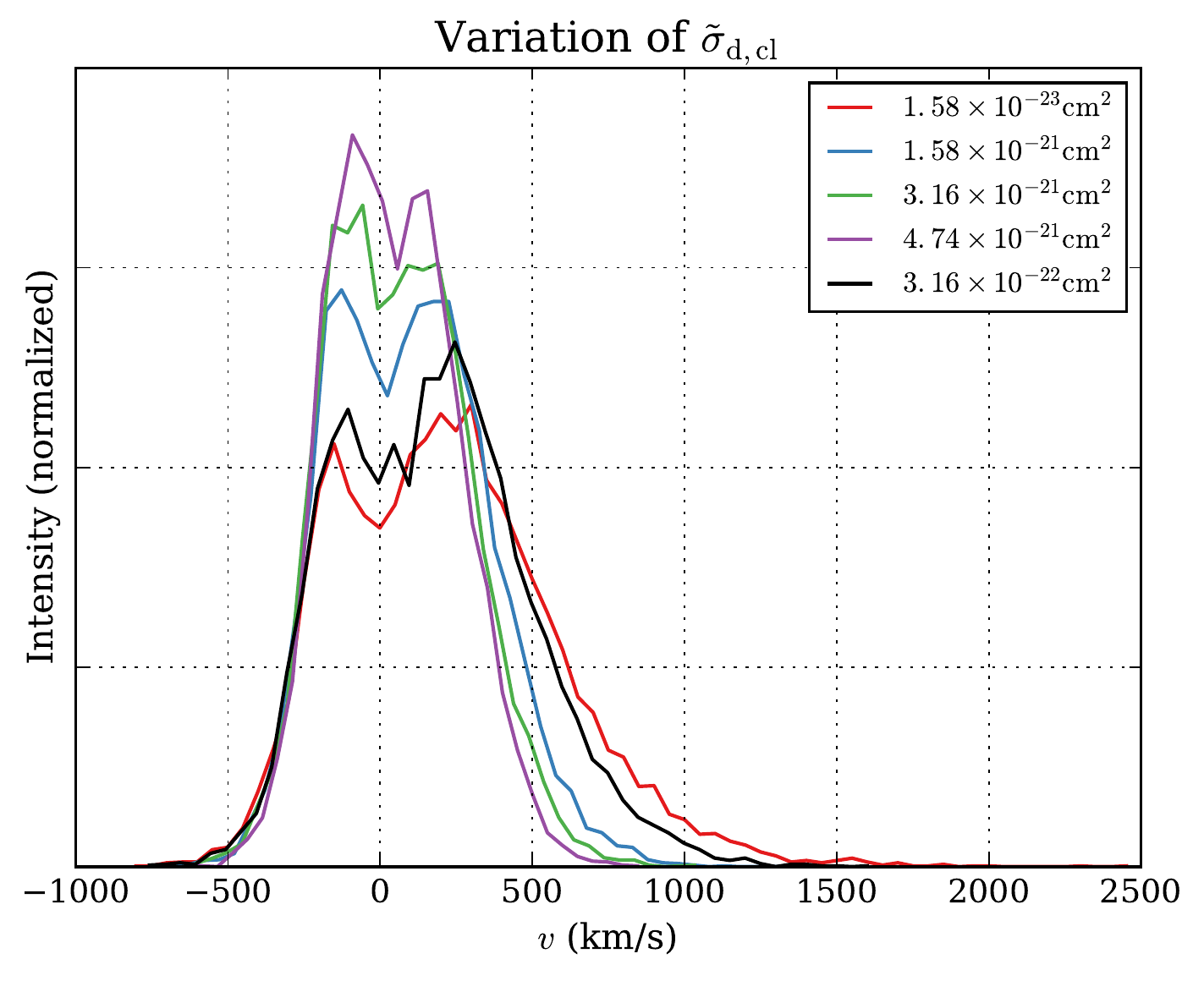}
  \caption{Variation of the dust number density in the ICM \textit{(left panel)} and in the clouds \textit{(right panel)}.}
  \label{fig:dust}
\end{figure}

\begin{figure}
\centering
\includegraphics[width=.33\textwidth]{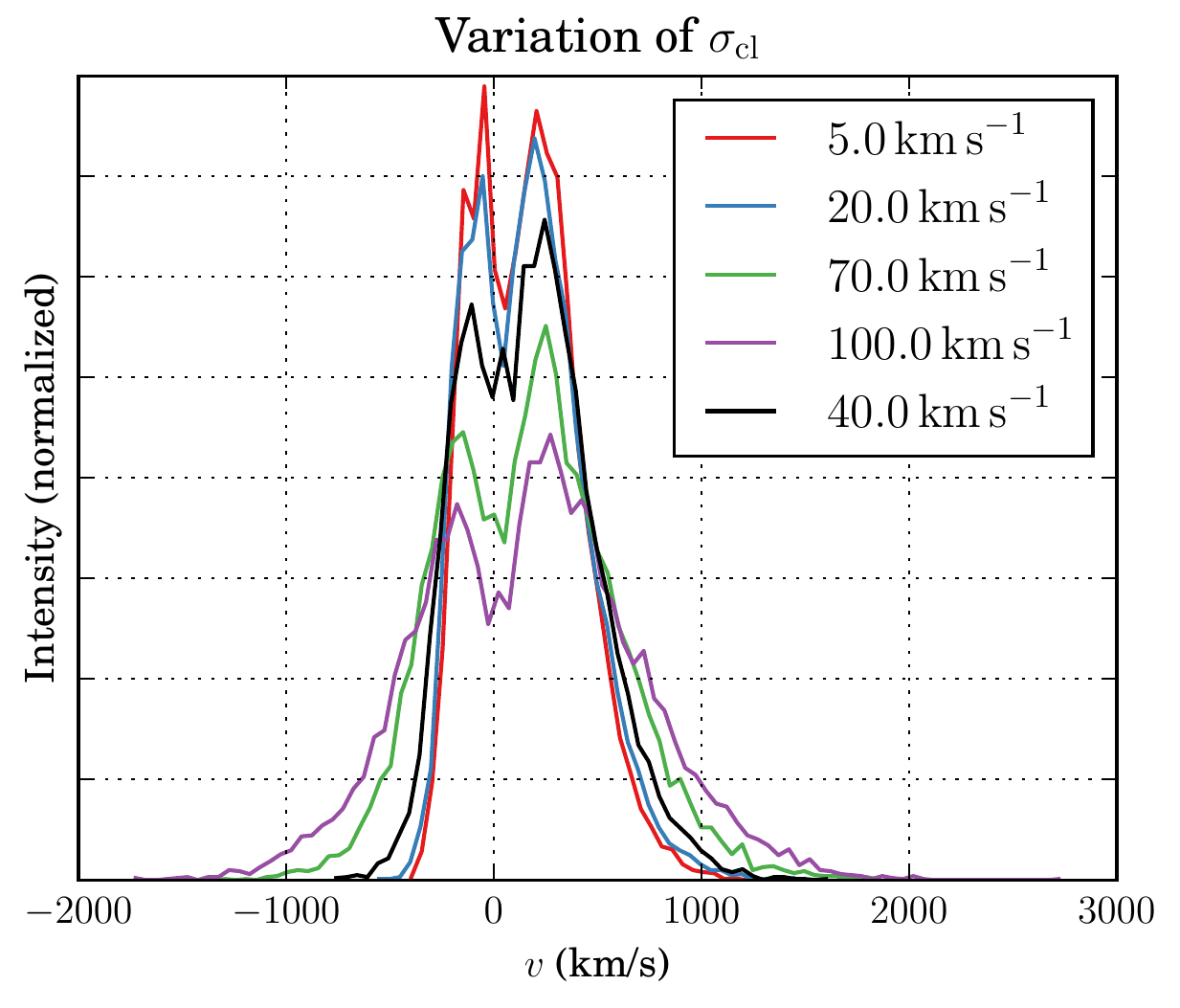}~
\includegraphics[width=.33\textwidth]{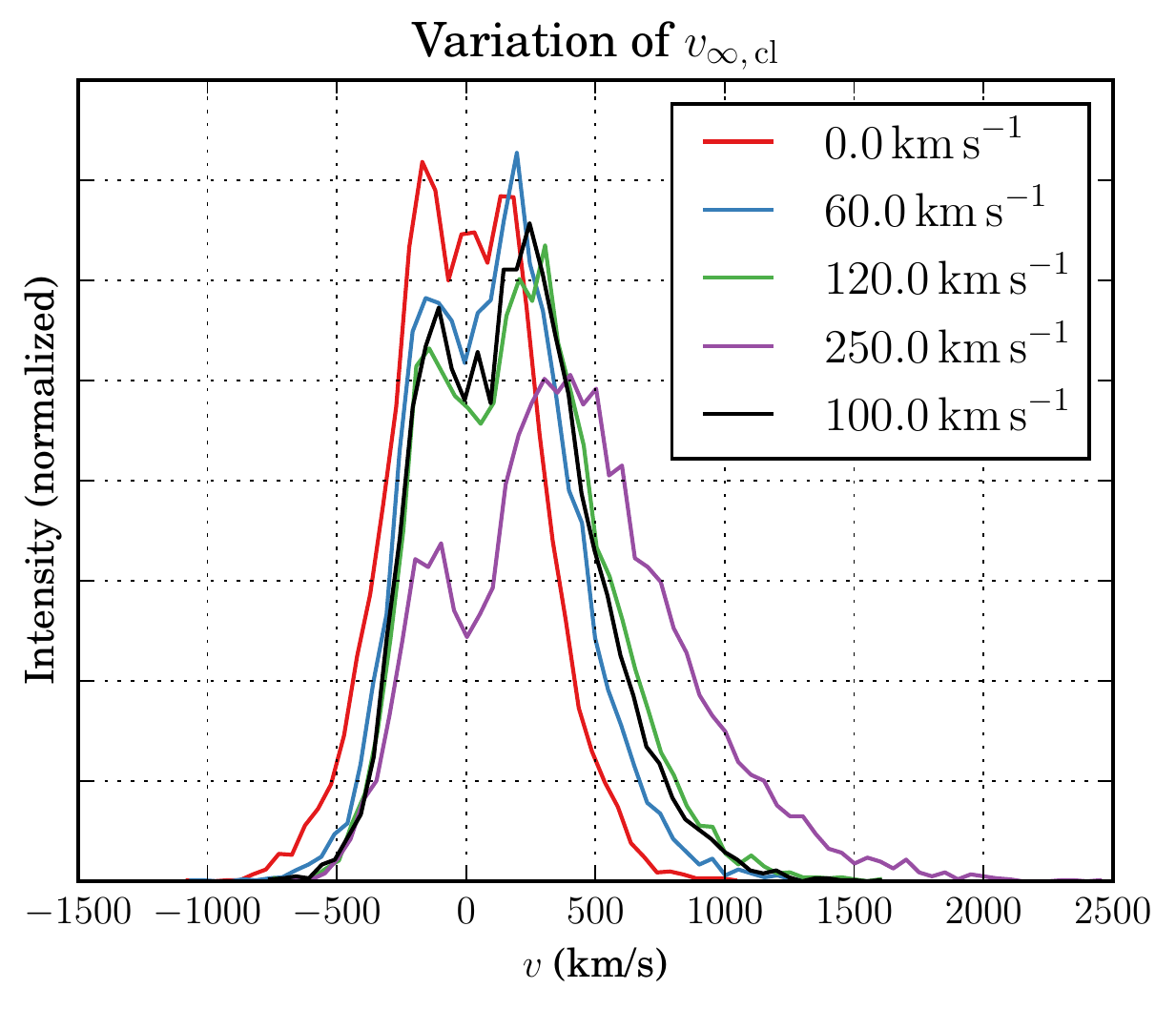}~
\includegraphics[width=.33\textwidth]{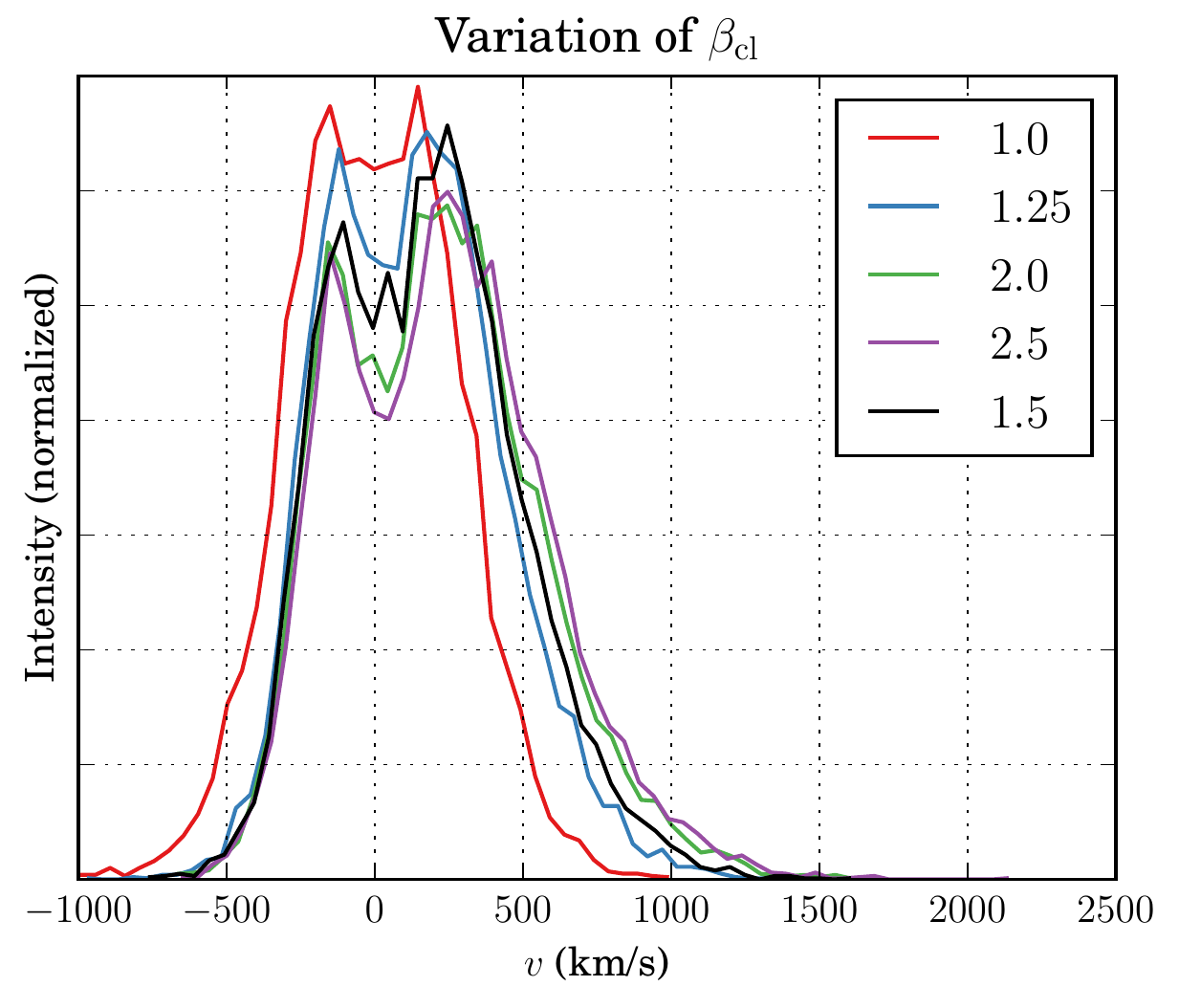}
\caption{Variation of the velocity field. In the \textit{left panel} the random component is altered, whereas in the \textit{central} panel the outflow speed is changed. The \textit{right panel} illustrates the impact of the change of $\beta_{\rm cl}$.}
  \label{fig:velocities}
\end{figure}

\begin{figure}
  \centering
  \plottwo{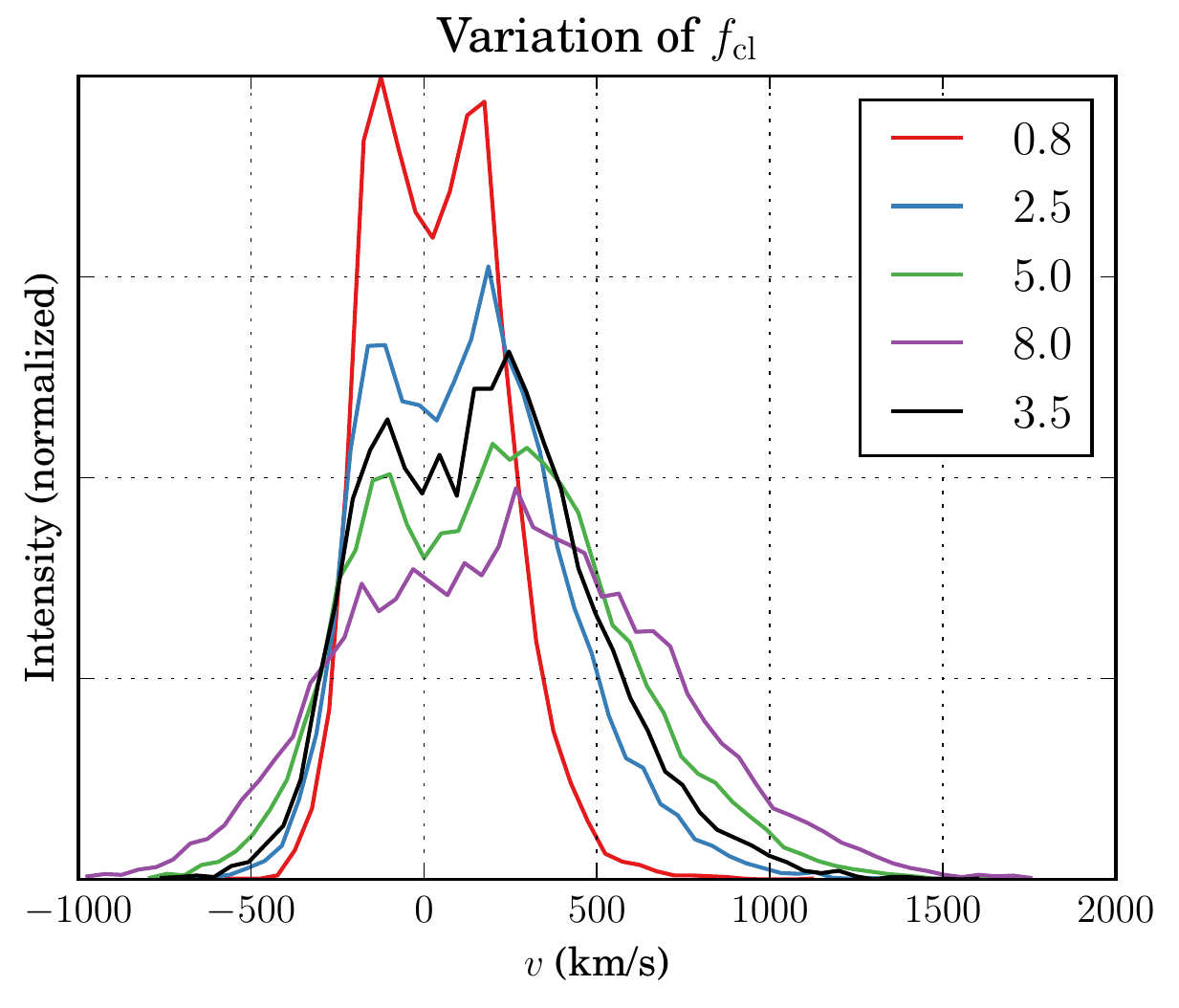}{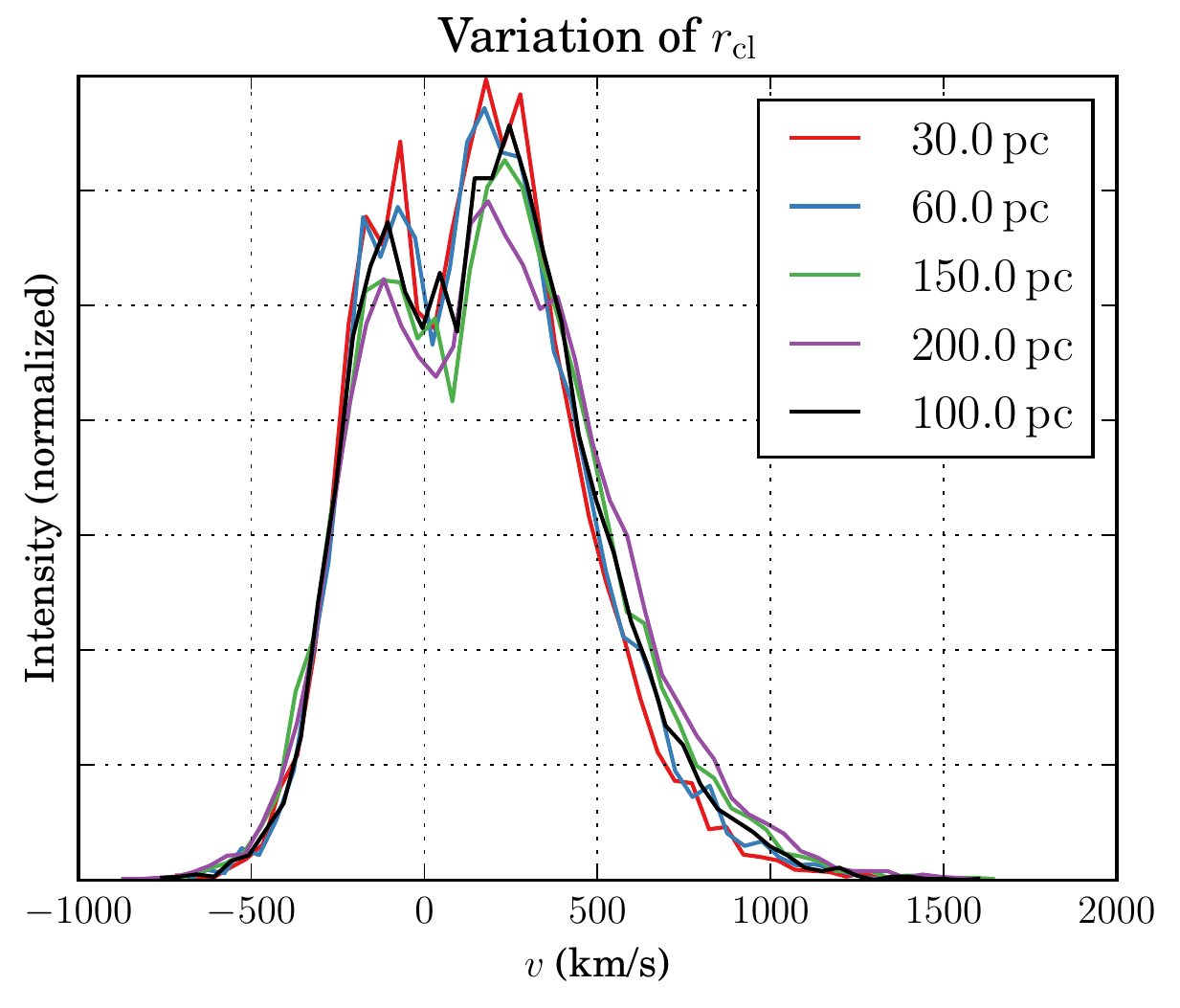}
  \caption{Variation of the geometrical setup given by the covering factor \textit{(left panel)}, and the cloud radius \textit{(right panel)}.}
  \label{fig:geometry}
\end{figure}

\begin{figure}
  \centering
\includegraphics[width=.33\textwidth]{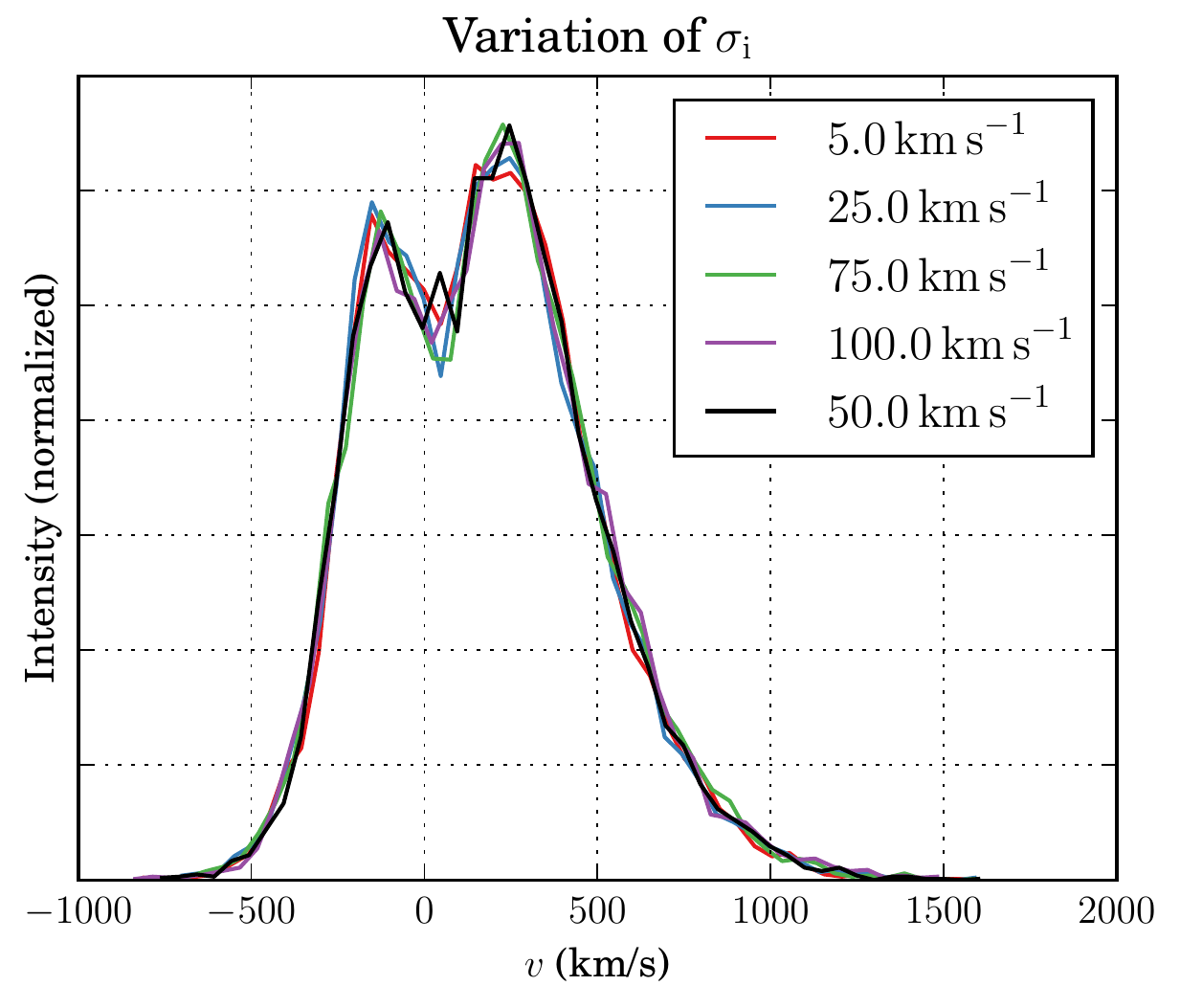}~
\includegraphics[width=.33\textwidth]{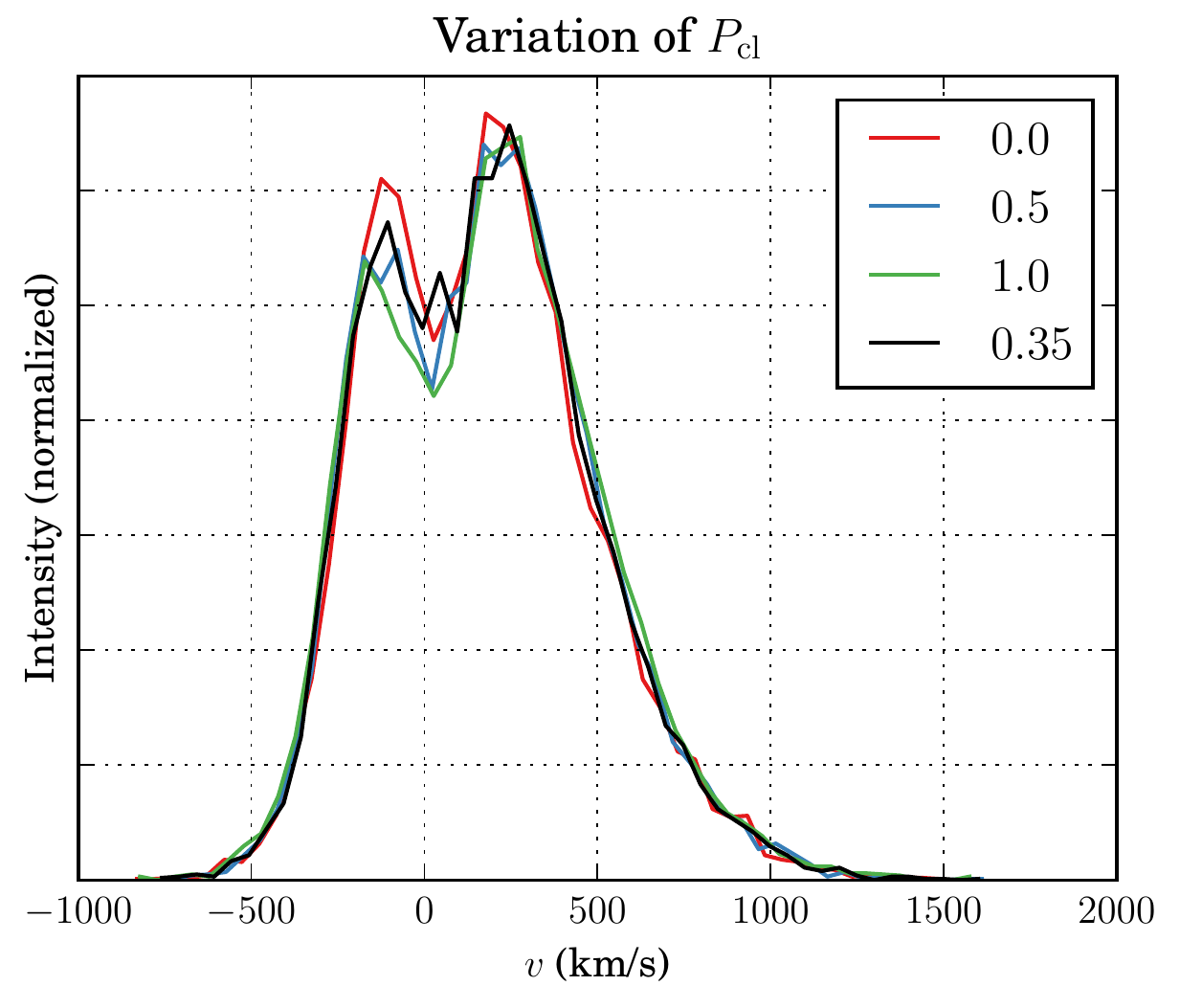}~
\includegraphics[width=.33\textwidth]{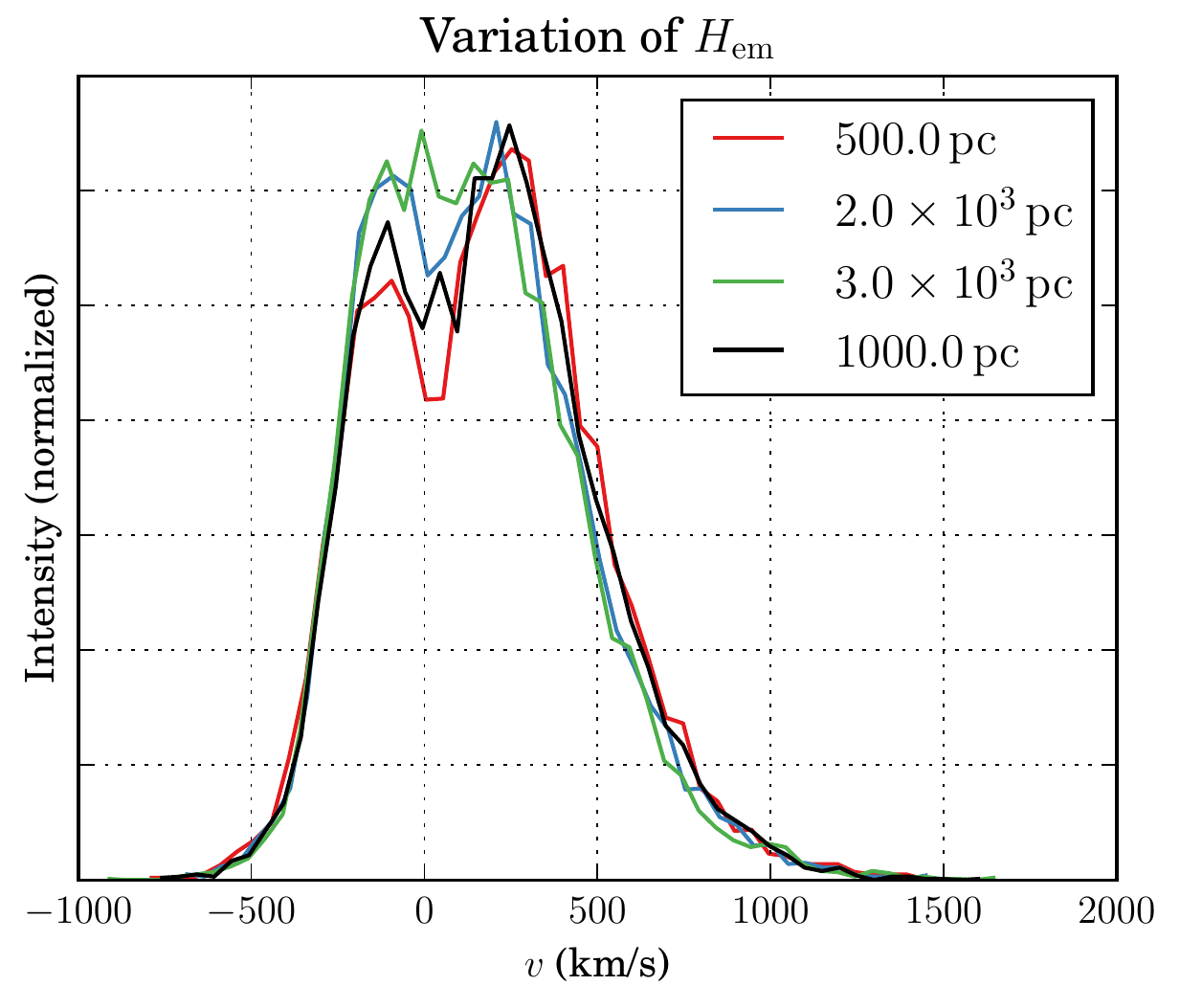}
  \caption{Impact of the emission properties on the emergent spectrum. In the \textit{left panel}, we varied the intrinsic width of the \Lya line, in the \textit{central panel} the probability to be emitted within a cloud, and in the \textit{right panel} the scale length of the emission site PDF.}
  \label{fig:emission}
\end{figure}

Starting with our fiducial set of parameters (see Table~\ref{tab:models}, which we display in Fig.~\ref{fig:temperatures}-\ref{fig:emission} as black solid line), we varied each clumpy model parameter individually. The results are as follows:
\begin{itemize}
\item Fig.~\ref{fig:temperatures} shows the impact of the ICM and cloud temperatures (left and right panel, respectively). Whereas the temperature of the ICM has a strong impact on the spectral shape, the cloud temperature seems to play a minor role. Generally speaking, an increase in $T_{\rm ICM}$ leads to a widening of the spectrum and a transition from a double to a single peaked spectrum.
\item We vary the hydrogen content of the ICM and in the clouds in Fig.~\ref{fig:hydrogen} (left \& right panel, respectively). Above a value of $n_{\HI, {\rm ICM}}\gtrsim 10^{-10}\ccm$ the ICM hydrogen number density affects the spectral shape. An greater value of $n_{\HI, {\rm ICM}}$ leads to a deepening of the central valley, and a increased peak separation. Contrary to that, $n_{\HI, {\rm cl}}$ does not seem so affect the spectral shape.
\item We show the impact of dust in Fig.~\ref{fig:dust}. In the left panel the dust content of the ICM was varied ($\zeta_d = \tilde\sigma_{\rm d, ICM} / \tilde\sigma_{\rm d, cl}$)-- which does not affect the spectral shape much. The dust content within the clumps, on the other hand, affects the spectrum much more strongly. With increasing dust optical depth, the spectrum becomes narrower and more symmetric, i.e., the red wing disappears. See \S\,\ref{sec:sens-lya-spectr} for a discussion of this effect.
\item Fig.~\ref{fig:velocities} shows the variation of the spectra due to the change in velocity parameters. All of them do affect the emergent spectrum. The increase in random motion ($\sigma_{\rm cl}$, left panel) results in a widening of the peaks (but not the peak separation). Having a stronger outflow ($v_{\infty,{\rm cl}}$, central panel) enhances the red and decreases the blue peak. It also widens the red peak. Similar effects can be observed when increasing $\beta_{\rm cl}$ (right panel).
\item Fig.~\ref{fig:geometry} illustrates the impact of changing $f_{\rm cl}$ (left panel) and the clouds' radii ($r_{\rm cl}$, right panel). Note, that in both cases the absolute number of clouds are varied as we made sure that the volume filling fraction of the clouds is kept constant when varying $r_{\rm cl}$. Altering the covering factor has a major impact on the spectral shape (widening of the peaks with increased $f_{\rm cl}$) whereas different values of $r_{\rm cl}$ (while keeping the volume filling factor constant) hardly change the spectra.
\item We display the spectra for different emission properties in Fig.~\ref{fig:emission}. None of them changes the spectrum significantly. We observe only a slight increase in the flux at line center for larger values of $H_{\rm em}$ (right panel).
\end{itemize}
Although changing parameters individually is illustrative, and the results are relatively easy to understand, we point out that these one-dimensional cuts through the $14$ dimensional parameter space does not capture the full complexity of the problem. We therefore caution the reader to over-interpret the results presented in this section.


\end{document}